\documentclass{emulateapj}
\usepackage{graphicx,amsfonts,natbib}
\shorttitle{GMCs in M33}
\shortauthors{Rosolowsky {\it et al.}}

\begin{document}
\title{Giant Molecular Clouds in M33\\ II -- High Resolution Observations}
\author{E. Rosolowsky\altaffilmark{1}, G. Engargiola, 
R. Plambeck and L. Blitz}
\affil{Department of Astronomy, University of California, Berkeley, CA 94720}
\altaffiltext{1}{eros@astro.berkeley.edu}
\begin{abstract}
We present $^{12}$CO$(1\to 0)$ observations of 45 giant molecular
clouds in M33 made with the BIMA array.  The observations have a
linear resolution of 20 pc, sufficient to measure the sizes of most
GMCs in the sample.  We place upper limits on the specific angular
momentum of the GMCs and find the observed values to be nearly an
order of magnitude below the values predicted from simple formation
mechanisms.  The velocity gradients across neighboring, high-mass GMCs
appear preferentially aligned on scales less than 500 pc.  If the
clouds are rotating, 40\% are counter-rotating with respect to the
galaxy.  GMCs require a braking mechanism if they form from the large
scale radial accumulation of gas.  These observations suggest that
molecular clouds form locally out of atomic gas with significant
braking by magnetic fields to dissipate the angular momentum imparted
by galactic shear.  The observed GMCs share basic properties with
those found in the Galaxy such as similar masses, sizes, and
linewidths as well as a constant surface density of 120 $M_\odot$
pc$^{-2}$.  The size--linewidth relationship follows $\Delta V \propto
r^{0.45\pm 0.02}$, consistent with that found in the Galaxy.  The
cloud virial masses imply that the CO-to-H$_2$ conversion factor has a
value of $2 \times 10^{20} \mbox{ H}_2 \mbox{ cm}^{-2}/(\mbox{K}
\mbox{ km} \mbox{ s}^{-1})$ and does not change significantly over the
disk of M33 despite a change of 0.8 dex in the metallicity.
\end{abstract}

\section{Introduction}
Many theories address molecular cloud formation, but few observations
can distinguish between them.  As a result, there is little consensus
on which process dominates \citep[][and references therein]{eg93}.
One avenue for testing these theories is to study molecular clouds in
a wide range of environments and to determine which aspects of the
environment set the cloud properties.  We have adopted this
approach in a study of the Local Group galaxy M33.  \citet[][Paper
I]{epb02} completed a survey of the entire optical disk of M33 in
$^{12}\mbox{CO}(J=1\to 0)$ with sufficient sensitivity to detect all
molecular clouds more massive than $1.5\times 10^5 M_{\odot}$.  This
survey represents the first {\it flux-limited} survey of giant
molecular clouds (GMCs) in any spiral galaxy.  While there are
complete surveys of molecular gas in the Milky Way, blending of
emission along the line of sight make it impossible to obtain a
complete catalog.  The only other flux-limited survey of a
molecular cloud population in another galaxy was completed by the
NANTEN group studying the Large Magellenic Cloud \citep{nanten}.
\citet{ws90} used the OVRO interferometer to study M33 GMCs at high
resolution (20 pc), but their observations were limited to clouds in
the central 1.5 kpc of the galaxy that are associated with optical
extinction and bright single-dish CO detections.

This paper is a high resolution follow-up to the survey of Paper I,
targeting fields known to contain bright GMCs over a large range of
galactic radii.  The $13''$ observations presented in Paper I have
sufficient linear resolution (50 pc) to resolve the emission into
molecular clouds, but not to measure the cloud sizes.  The higher
resolution observations presented here (20 pc) resolve most of the
GMCs, and provide additional information about their masses, angular
momenta and morphologies.  These data show that the GMCs in M33 are
similar to those found in the Milky Way.  In addition, the data
constrain cloud formation theories, and they suggest that the
CO-to-H$_2$ conversion factor is constant across the galaxy.

\section{Observations}

\label{obs}
\subsection{Interferometer Observations}

\begin{figure}[ht]
\begin{center}
\plotone{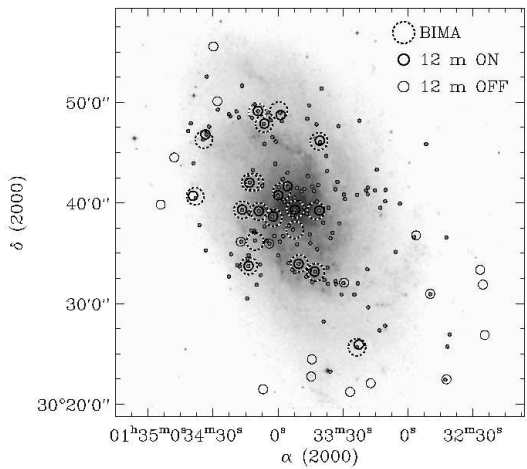} \figcaption{\label{fields} The locations of the 20
fields observed by the BIMA interferometer and the 18 locations
observed by the UASO 12-m overlaid on smoothed $I$ band DSS image of
the galaxy.  The locations of the catalog sources from Paper I are
plotted as grey circles and were used to target the C-array and 12-m
observations.}
\end{center}
\end{figure}

We observed 17 fields in $^{12}$CO $(J=1\to 0)$ using the BIMA array
\citep{bima} in Fall 2000 and Spring 2001.  We chose fields that
contain the highest mass GMCs cataloged in Paper I (Figure
\ref{fields}).  The observations were made in the C-array
configuration yielding a synthesized beam FWHM of $\sim 6''$.  We
observed the CO line in the upper sideband with a velocity resolution
of 1.016 km s$^{-1}$.  System temperatures typically ranged between
420 K and 730 K.  Table \ref{table1} lists the field center,
synthesized beam size, integration time, and rms noise for each field.
We tried to obtain $< 0.2$ Jy beam$^{-1}$ rms noise in each field
while maintaining good $UV$ coverage.  Three additional fields (Fields
18 through 20 in Table \ref{table1}) were observed in conjunction with
the survey presented in Paper I and are discussed in more detail in
that paper.  Because of the longer integration times and smaller
synthesized beam sizes, the high resolution observations can detect
lower mass clouds than the survey in Paper I.

\begin{table}[ht]
\begin{center}
{\scriptsize
\caption{Observations of 20 fields in M33 using the BIMA
millimeter interferometer. \label{table1}}
\begin{tabular}{cccccc}
\tableline\tableline
Field & $\alpha$ & $\delta$ &
Beam Size & Source Time & $\sigma$
\tablenotemark{a} \\
Name & (2000) & (2000) & 
(\arcsec) & (hours) & (Jy beam$^{-1}$) \\

\tableline
1 & $1^{\mbox{\scriptsize h}}\ 34^{\mbox{\scriptsize m}}\ $ 12\fs 7 & 
30\degr\ 42\arcmin\ 2\farcs 80 & 
5.3 $\times$ 5.1 & 
8.22
 & 
0.127 \\
2 & $1^{\mbox{\scriptsize h}}\ 34^{\mbox{\scriptsize m}}\ $ 8\fs 76 & 
30\degr\ 39\arcmin\ 8\farcs 78 & 
5.8 $\times$ 5.4 & 
0.872
 & 
0.199
 \\
3 & $1^{\mbox{\scriptsize h}}\ 34^{\mbox{\scriptsize m}}\ $ 6\fs 57 & 
30\degr\ 47\arcmin\ 50\farcs 8 & 
5.7 $\times$ 5.4 & 
0.872
 & 
0.200
 \\
4 & $1^{\mbox{\scriptsize h}}\ 33^{\mbox{\scriptsize m}}\ $ 23\fs 5 & 
30\degr\ 25\arcmin\ 39\farcs 8 & 
5.8 $\times$ 4.9 & 
0.255
 & 
0.205
 \\
5 & $1^{\mbox{\scriptsize h}}\ 33^{\mbox{\scriptsize m}}\ $ 41\fs 0 & 
30\degr\ 46\arcmin\ 7\farcs 43 & 
5.8 $\times$ 5.2 & 
1.34
 & 
0.159
 \\
6 & $1^{\mbox{\scriptsize h}}\ 34^{\mbox{\scriptsize m}}\ $ 38\fs 2 & 
30\degr\ 40\arcmin\ 38\farcs 1 & 
5.8 $\times$ 5.2 & 
1.34
 & 
0.160
 \\
7 & $1^{\mbox{\scriptsize h}}\ 33^{\mbox{\scriptsize m}}\ $ 50\fs 1 & 
30\degr\ 33\arcmin\ 58\farcs 3 & 
5.8 $\times$ 5.2 & 
1.34
 & 
0.161
 \\
8 & $1^{\mbox{\scriptsize h}}\ 34^{\mbox{\scriptsize m}}\ $ 10\fs 7 & 
30\degr\ 42\arcmin\ 2\farcs 80 & 
6.4 $\times$ 5.4 & 
1.88
 & 
0.178
 \\
9 & $1^{\mbox{\scriptsize h}}\ 33^{\mbox{\scriptsize m}}\ $ 42\fs 5 & 
30\degr\ 33\arcmin\ 6\farcs 61 & 
5.7 $\times$ 5.0 & 
4.44
 & 
0.0912
 \\
10 & $1^{\mbox{\scriptsize h}}\ 34^{\mbox{\scriptsize m}}\ $ 2\fs 20 & 
30\degr\ 38\arcmin\ 31\farcs 1 & 
6.9 $\times$ 4.9 & 
2.34
 & 
0.212
 \\
11 & $1^{\mbox{\scriptsize h}}\ 34^{\mbox{\scriptsize m}}\ $ 13\fs 5 & 
30\degr\ 33\arcmin\ 44\farcs 2 & 
6.5 $\times$ 6.0 & 
1.33
 & 
0.262
 \\
12 & $1^{\mbox{\scriptsize h}}\ 33^{\mbox{\scriptsize m}}\ $ 51\fs 9 & 
30\degr\ 39\arcmin\ 22\farcs 8 & 
6.5 $\times$ 6.0 & 
1.33
 & 
0.261
 \\
13 & $1^{\mbox{\scriptsize h}}\ 34^{\mbox{\scriptsize m}}\ $ 34\fs 2 & 
30\degr\ 46\arcmin\ 16\farcs 7 & 
6.6 $\times$ 6.3 & 
1.45
 & 
0.221
 \\
14 & $1^{\mbox{\scriptsize h}}\ 33^{\mbox{\scriptsize m}}\ $ 59\fs 6 & 
30\degr\ 49\arcmin\ 10\farcs 8 & 
6.5 $\times$ 6.0 & 
2.20
 & 
0.213
 \\
15 & $1^{\mbox{\scriptsize h}}\ 34^{\mbox{\scriptsize m}}\ $ 16\fs 4 & 
30\degr\ 39\arcmin\ 18\farcs 1 & 
6.7 $\times$ 6.2 & 
1.92
 & 
0.196
 \\
16 & $1^{\mbox{\scriptsize h}}\ 34^{\mbox{\scriptsize m}}\ $ 10\fs 5 & 
30\degr\ 36\arcmin\ 10\farcs 0 & 
6.5 $\times$ 5.8 & 
3.09
 & 
0.234
 \\
17 & $1^{\mbox{\scriptsize h}}\ 34^{\mbox{\scriptsize m}}\ $ 9\fs 50 & 
30\degr\ 49\arcmin\ 1\farcs 20 & 
6.7 $\times$ 6.1 & 
16.0
 & 
0.0832
\\
18 \tablenotemark{b} & $1^{\mbox{\scriptsize h}}\ 33^{\mbox{\scriptsize m}}\ $ 58\fs 3 & 
30\degr\ 41\arcmin\ 6\farcs 99 & 
6.6 $\times$ 6.6 & 
4.00
 & 
0.155
 \\
19 \tablenotemark{b}& $1^{\mbox{\scriptsize h}}\ 33^{\mbox{\scriptsize m}}\ $ 42\fs 5 & 
30\degr\ 39\arcmin\ 13\farcs 0 & 
6.6 $\times$ 6.5 & 
4.00
 & 
0.153
 \\
20 \tablenotemark{b} & $1^{\mbox{\scriptsize h}}\ 33^{\mbox{\scriptsize m}}\ $ 52\fs 2 & 
30\degr\ 37\arcmin\ 19\farcs 0 & 
6.6 $\times$ 6.5 & 
4.00
 & 
0.155
 \\
\tableline
\end{tabular}
\tablenotetext{a}{For a 2 km s$^{-1}$ channel.}
\tablenotetext{b}{These observations of these fields are discussed in
Paper I.}
}
\end{center}
\end{table}

Our data reduction followed the algorithm developed for the BIMA
Survey of Nearby Galaxies \citep{song}.  Only the C-array data were
used in producing the final data cubes because of their significantly
lower noise levels.  Observations of Uranus (Fall 2000) and Mars
(Spring 2001) were used to establish the flux of the phase calibrator
0136+478 (3.6 Jy in Fall 2000, 2.7 Jy in Spring 2001).  Bootstrapping
the quasar flux to the M33 data established the flux calibration.
Binning velocity channels set the final velocity resolution at 2.00 km
s$^{-1}$.  We used uniform weighting of the $UV$ data in the
inversion, choosing a grid spacing that resulted in $1.5''$ pixels for the
final maps, and we cleaned the cube using a H\"ogbom algorithm after
correcting for the primary beam attenuation.

\subsection{Single Dish Observations}

Since interferometers filter out zero-spacing flux from the observed
object, complete flux recovery requires additional observations using
a single dish telescope.  To measure the total flux from the GMC
fields, we observed 18 fields over the course of three nights
using the University of Arizona Steward Observatory 12-m telescope at
Kitt Peak.  The selected fields for the 12-m observations are
displayed in Figure \ref{fields}.  To check for faint emission, each
field was observed to attain a signal-to-noise ratio at the peak of
the CO line of at least 10.  Every two hours, Jupiter or Saturn was
observed to optimize the pointing and focus of the telescope.  During
sunset, the pointing and focus checks were done before observing each
new field to compensate for rapid variations in the thermal stresses
on the telescope.  Both polarizations were observed with the 500-kHz
and 1-MHz filter banks in parallel mode.  The median system
temperature was 350 K for the observations and the median column of
water was 0.7 mm.

The observations used absolute off positions which targeted
low-significance detections in the D-array survey (see Paper I).
These observations help to establish the completeness limit in the
survey by determining the fraction of real detections of low
significance.  The positions of the off fields were selected to be
well separated in velocity from the strong molecular cloud emission in
the on position.  Emission in the off fields appears as absorption
features in the resulting spectra.  Only two of the off fields
contained emission at the expected velocities.

We reduced the data by flagging bad channels in every spectrum,
combining the polarizations to reduce the noise, and fitting a
quadratic baseline.  The vane calibration of the antenna temperature
was found to be accurate to better than 10\% using continuum
observations of Jupiter and line observations of DR21(OH).  All
spectra were observed to an rms noise level of at most 10 mK in a 2.6
km s$^{-1}$ channel, equivalent to a molecular gas surface density of
of 0.1 $M_{\odot}$ pc$^{-2}$.

Finally, we observed a 5 $\times$ 5 pointing mosaic centered on the
most massive cloud complex in the galaxy (Field 17 in Table
\ref{table1}) to map all flux associated with this cloud.  The single
dish map was combined with both the C and D array BIMA data using
MIRIAD's {\sf imcomb} routine.  The difference between the
fully-recovered data cube and the cube generated solely from the
interferometer data shows the location and spatial structure of the
emission that is detected only by the 12-m telescope.  

\subsection{Flux Recovery}
\label{fluxrec}

By comparing the interferometer and single dish data, we estimate the
amount of flux that has been resolved out by the interferometer or not
selected by our cloud identification techniques (\S \ref{masking}).
We compare each of the 18 single dish spectra to a representative
spectrum generated from the interferometer data by averaging together
the spectra in the data cube and correcting for the beam profile of
the 12-m.  To improve the signal to noise in the interferometer
spectrum, only pixels with significant ($> 2\sigma$) CO emssion are
included in the average.  Both the C-array observations and the
D-array observations discussed in Paper I recover less than 100\% of
the flux as expected.  The median ratios of the interferometer and
single dish integrated intensities ($L_{\mathrm{CO}}$) are:
\begin{eqnarray*}
\langle L_{\mathrm{CO,C}}/L_{\mathrm{CO,12-m}}\rangle
&=&0.53^{+0.22}_{-0.09} \mbox{ and } \\
\langle
L_{\mathrm{CO,D}}/L_{\mathrm{CO,12-m}}\rangle& =&0.62^{+0.30}_{-0.13}.
\end{eqnarray*}
Because the distribution of this ratio is asymmetric about its median,
we quote the positive and negative dispersions separately.  The 12-m
observations indicate that a significant fraction of emission is
resolved out or masked out (see \S \ref{masking}).  However, comparing
the peak antenna temperatures in the spectra shows that:
\begin{eqnarray*}
\langle T_{A,\mathrm{C}}/T_{A,\mathrm{12-m}}\rangle
&=&0.77^{+0.22}_{-0.17} \mbox{ and } \\ \langle
T_{A,\mathrm{D}}/T_{A,\mathrm{12-m}}\rangle &=&0.89^{+0.11}_{-0.13},
\end{eqnarray*} 
while the widths of the interferometer lines are narrower:
\begin{eqnarray*}
\langle \Delta v_{\mathrm{CO,C}} / \Delta v_{\mathrm{CO,12-m}}\rangle
&=&0.71_{-0.12}^{+0.22}\mbox{ and }\\
 \langle \Delta v_{\mathrm{CO,D}} / \Delta
v_{\mathrm{CO,12-m}}\rangle &=& 0.72^{+0.25}_{-0.16}.
\end{eqnarray*}

Three effects can lead to a discrepancy between the interferometer and
single dish data: (1) calibration errors, (2) extended CO emission
resolved out by the interferometer, and (3) limited sensitivity or
masking effects in the interferometer data.

Calibration discrepancies would appear as a relative scaling of the
observed spectra by a constant factor.  However, the ratios of the
peak temperatures and linewidths indicate better flux recovery in the
middle of the line.  We conclude that scaling alone cannot not be the
source of flux loss.

Observations of the off positions using the 12-m telescope indicate
that there is no extended CO emission on kiloparsec scales with
sufficient surface brightness to account for the discrepancy between
the interferometer and the single dish observations.  Each of the 18
single dish spectra have a $3\sigma$ noise level of 0.07 K km s$^{-1}$
(0.3 $M_{\odot}$ pc$^{-2}$); however, accounting for the observed
difference requires a surface brightness of 1.0 K km s$^{-1}$ (4.3
$M_{\odot}$ pc$^{-2}$).  To search for even fainter emission, we
averaged together the 16 empty off positions, shifting the velocities
to align any faint emission at the \ion{H}{1} velocity. No line is
present in the resulting spectrum, which has a $3\sigma$ surface
brightness sensitivity of 0.019 K km s$^{-1}$ ($0.08\ M_{\odot}$
pc$^{-2}$).  Since the off positions are separated by projected
distances of roughly a kiloparsec, we conclude that a CO emission
component on this scale can account for no more than 2\% of the
discrepancy between the interferometer and the single dish fluxes.

\begin{figure}
\begin{center}
\plotone{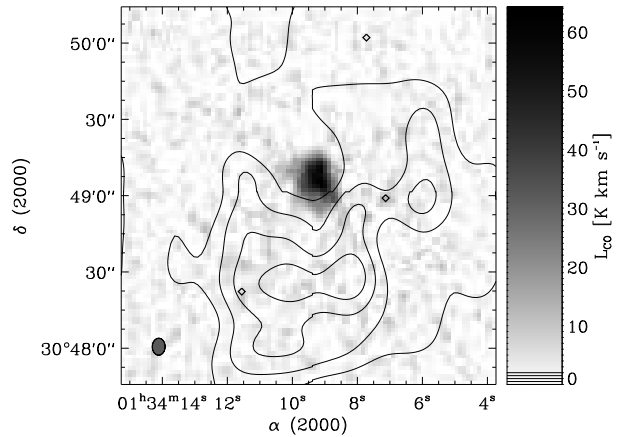} \figcaption{\label{diffmap} Map of emission in
Field 17 of the interferometer observations.  The greyscale shows the
map generated by combining the interferometer and single-spacing data.
The contours show the distribution of flux present in the combined map
but not in the interferometer map.  The contours are spaced at 0.5 K
km s$^{-1}$ starting at 0.5 K km s$^{-1}$, approximately $1\sigma$ for
this map. The relative levels of the contours are indicated as lines
on the scale bar, showing the diffuse emission is of low
surface brightness.  The three diamonds overplotted are the locations
of additional clouds found in deep C-array integrations.}
\end{center}
\end{figure}

The extra emission detected in the single-dish spectra is likely from
low mass clouds surrounding each GMC lying below our clipping
threshold.  Figure \ref{diffmap} compares the map of Field 17
generated from interferometer data alone with the map generated by
merging interferometer data with the single dish map.  The greyscale
image is the fully recovered map and the contours represent the {\it
additional} emission seen only in the fully recovered map.  The
relative levels of the contours to the greyscale are indicated at the
bottom of the scale bar, showing the additional flux recovered is a
very low surface brightness component.  Taking the Fourier transform
of the added emission shows that $> 80\%$ of the added flux would be
detected by the interferometer with more integration time.  The
spatial filtering of the interferometer occurs only at $UV$ distances
of $< 5 \mbox{k}\lambda$ and does not account for the discrepancy
between the two telescopes.  Most of the added emission is excluded
from the interferometer map simply by virtue of being below the
significance threshold for our masking methods.  Also plotted in
Figure \ref{diffmap} are the locations of low mass clouds seen only in
the C-array data, where the synthesized beam is well matched to the
cloud sizes.  Two of these clouds appear in the distribution of
diffuse emission suggesting clumpy structure within this envelope.
The diffuse emission seems principally associated with the GMCs, and a
discussion of what this observation implies for cloud formation
appears in \S\ref{finis}. We conclude that the interferometer
observations detect the central clouds of complexes in the galaxy
containing high mass GMCs ($M > 1.5 \times 10^5 M_{\odot}$, Paper I).

\subsection{Cloud Identification}
\label{masking}
We defined clouds in the position-velocity interferometer data cubes
as connected regions in three dimensions containing more than 50
pixels, each with a value greater than $2\sigma_{rms}$.  Our criteria
guarantee that {\it at least} five statistically independent regions
are included in each cloud because interferometer data cubes are
oversampled by a factor of 3 in each spatial direction, though the
velocity channels are nearly independent.  For five joint $2\sigma$
detections, the likelihood of false detection is $5\times 10^{-4}$ in
these data cubes which have $\sim 10^5$ independent elements.
Analysis of the noise in the data indicates that the distribution is
sufficiently Gaussian to justify such a treatment.  We simulated
$10^3\ UV$ data sets mimicking the observation conditions, correlator
configuration, and baselines of the array.  The cloud extraction
program found no false detections for these selection criteria.

\begin{figure*}
\begin{center}
\plotone{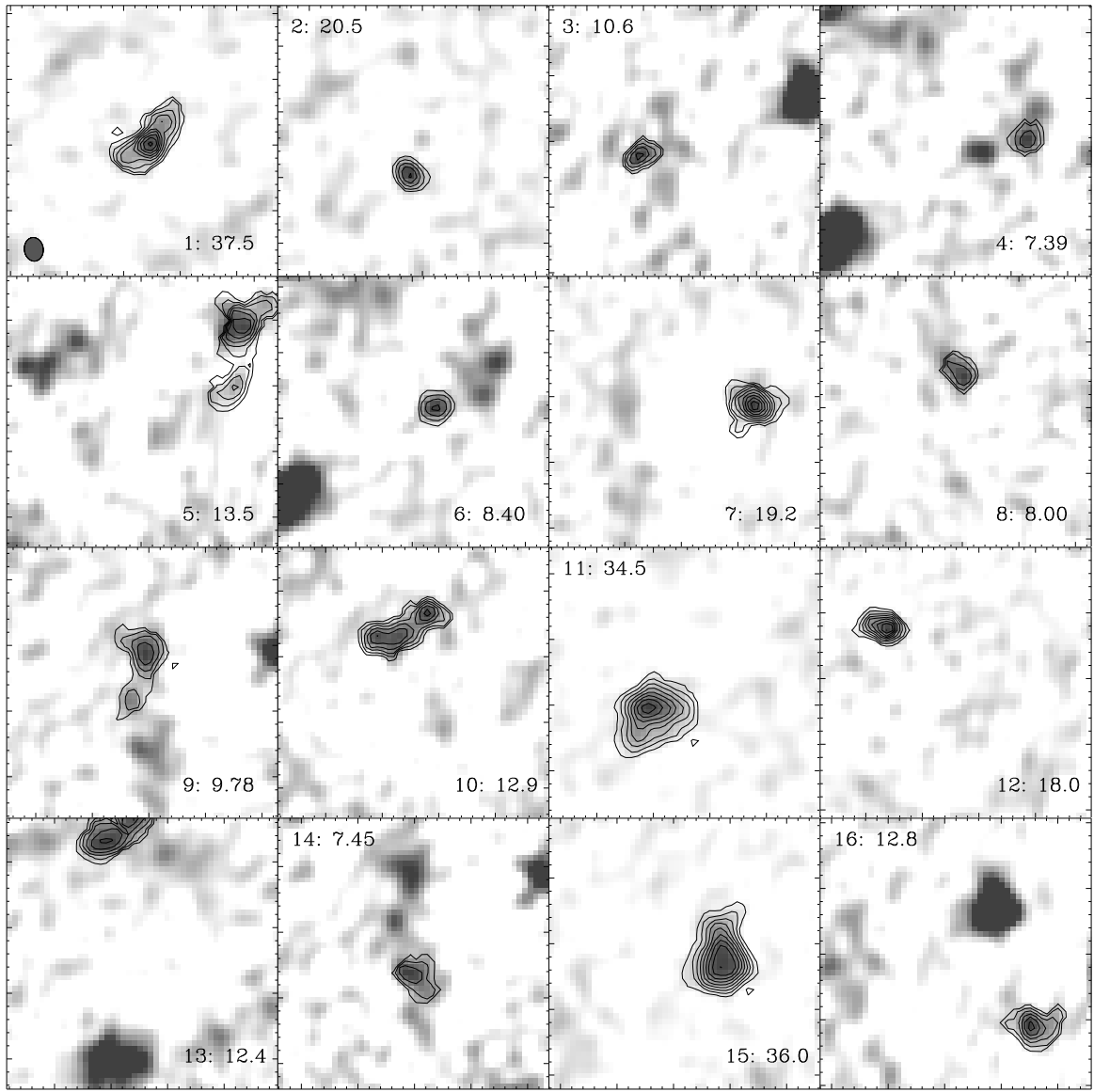} \figcaption{Maps of individual molecular clouds in
M33.  The greyscale map is the CO brightness, integrated
over all velocities contributing to the cloud.  Each box is 300 pc
square. The maps are annotated in the corner with
``$N$:$L_{\mathrm{CO}}$'' where $N$ is the number of the cloud in
Table \ref{cloudprop} and $L_{\mathrm{CO}}$ is the peak integrated
intensity (K km s$^{-1}$) in the cloud which sets the top of the
greyscale.  The bottom of the greyscale is set at 0 K km s$^{-1}$.
Other clouds in the same field of view with higher peak intensities
appear saturated.  The typical beam is plotted in the first panel of
the figure.  Plotted over the greyscale are contours of integrated
intensity for the cloud under consideration.  The spacing between the
contours is 2 K km s$^{-1}$ ($\sim 2\sigma_{rms}$) if the peak
integrated intensity ($L_{\mathrm{CO}}$) is less than 20 K km s$^{-1}$
and 4 K km s$^{-1}$ otherwise.  The maps show a wide variety of
morphologies and brightnesses.
\label{maps}}
\end{center}
\end{figure*}

This algorithm extracts 45 molecular clouds from the data cubes,
including all clouds in these fields cataloged in Paper I.  To account
for low amplitude signal at the edges of the clouds, the regions were
expanded by 1 pixel in both the position and the velocity directions.
Figure \ref{maps} shows maps of the individual clouds.  Five of the 45
clouds are not completely recovered since they lie near the half power
point of the primary beam.  Because their peaks are included in the
maps, we estimate that the truncation will not increase their masses
and areas by more than a factor of 2.

\begin{figure*}
\begin{center}
\figurenum{\ref{maps}}
\plotone{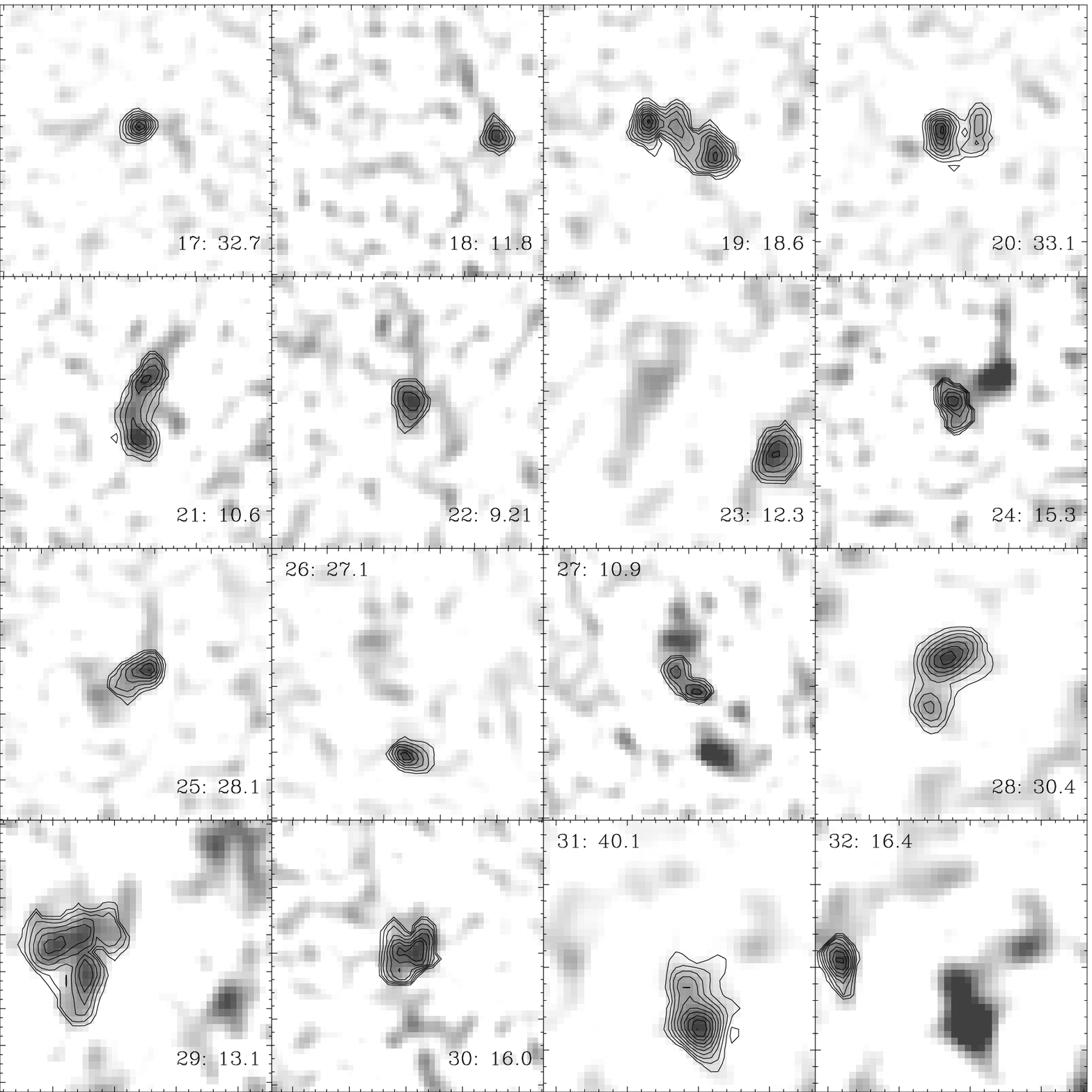}
\figcaption{{\it Continued}}
\end{center}
\end{figure*}

\begin{figure*}
\begin{center}
\figurenum{\ref{maps}}
\plotone{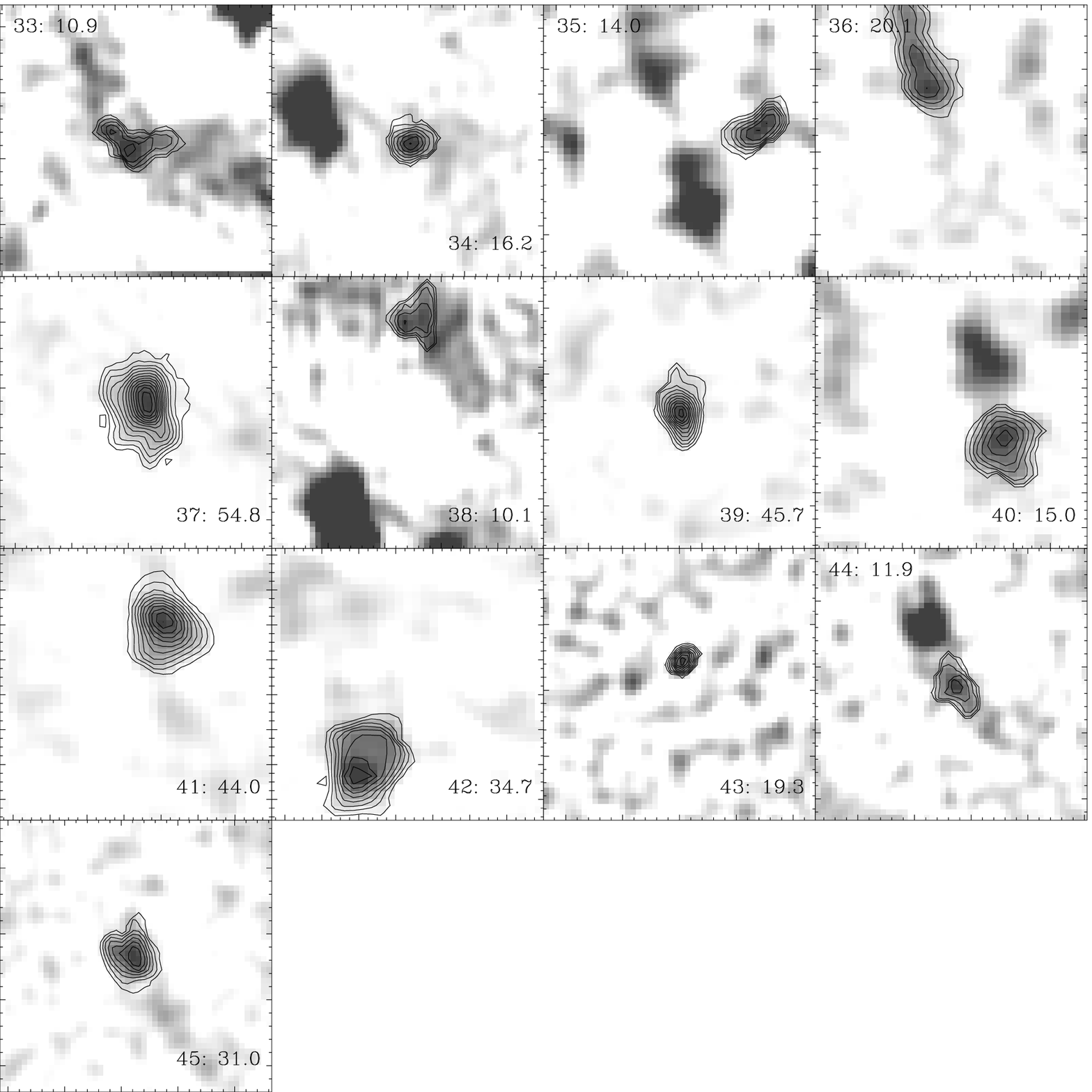}
\figcaption{{\it Continued}}
\end{center}
\end{figure*}

%
%

\section{GMC Properties}
\label{cloudprops}
The observed parameters of the 45 clouds are listed in Table
\ref{cloudprop}. The analysis uses a distance to M33 of 850 kpc
\citep{k98}.

\begin{table*}
\begin{center}
{\scriptsize
\caption{\label{cloudprop}Cloud properties for clouds found in M33}
\begin{tabular}{ccccccccccccccccc}
\tableline\tableline
{Cloud} & 
{$(\alpha,\delta)$\tablenotemark{a}} &
{Alternative\tablenotemark{b}} & {V$_{\mbox{LSR}}$} &
{$V_{\mbox{FWHM}}$} & {$r$\tablenotemark{c}} &
{$M_{\mbox{VT}}$\tablenotemark{c}} & 
{$M_{\mbox{CO}}$\tablenotemark{c}} &
{$|\nabla v|$\tablenotemark{c}} & 
{$\phi_{\nabla}$\tablenotemark{c}} &
{$j_1$\tablenotemark{c}} & {$j_2$}\\
{Number} & 
{$('','')$} &
{Names} & {(km s$^{-1}$)} &
{(km s$^{-1}$)} & {(pc)} &
{($10^4$ M$_{\odot}$)} & {($10^4$ M$_{\odot}$)} &
{(0.01 km s$^{-1}$ pc$^{-1}$)} & 
{($^\circ$)} 
& {pc km s$^{-1}$} & {pc km s$^{-1}$}\\	
\tableline
1 & $19,-18$ & EPRB19,WS1,2,4,5 & $-168.$ & 7.7 & 27. & 31. & 26. & 8.38 & $178$ & 19.2 & 25.3\\
2 & $29,-36$ & EPRB132,WS3 & $-149.$ & 5.5 & --- & --- & 6.7 & 2.29 & $-64$ & --- & 2.14\\
3 & $-82,-41$ & WS33,34 & $-162.$ & 6.7 & --- & --- & 3.6 & 9.58 & $-82$ & --- & 6.18\\
4 & $71,100$ & EPRB31,WS29 & $-215.$ & 6.1 & 22. & 16.\tablenotemark{d} & 4.2 & 10.1 & $-44$ & 15.3 & 47.4\\
5 & $61,119$ & EPRB31,WS29 & $-223.$ & 11. & 48. & 100\tablenotemark{d} & 14. & 13.6 & $-60$ & 93.9 & 103.\\
6 & $81,92$ & EPRB30,WS30 & $-214.$ & 4.9 & --- & --- & 2.9 & 11.9 & $19$ & --- & 10.3\\
7 & $-9,-129$ & WS10 & $-151.$ & 8.4 & 13. & 17. & 9.9 & 8.36 & $143$ & 4.00 & 13.4\\
8 & $26,-117$ & EPRB75,WS6 & $-158.$ & 4.9 & --- & --- & 2.9 & 3.49 & $-33$ & --- & 2.44\\
9 & $27,-129$ & EPRB75,WS8,9 & $-152.$ & 4.4 & 13. & 4.6 & 5.3 & 2.20 & $-9$ & 1.06 & 4.22\\
10 & $113,115$ & EPRB18,WS27,28 & $-221.$ & 6.5 & 33. & 27.\tablenotemark{d} & 12. & 1.90 & $-16$ & 6.43 & 8.87\\
11 & $119,69$ & EPRB3,WS21,22,23(?) & $-210.$ & 9.6 & 32. & 56. & 41. & 9.53 & $24$ & 30.0 & 46.0\\
12 & $52,-113$ & EPRB16 & $-168.$ & 7.6 & --- & --- & 6.7 & 10.6 & $32$ & --- & 12.8\\
13 & $118,133$ &  & $-208.$ & 7.2 & 22. & 22. & 11. & 3.29 & $-46$ & 5.06 & 4.56\\
14 & $25,-157$ & EPRB75,WS7 & $-153.$ & 4.2 & 6.7 & 2.3 & 3.7 & 7.70 & $-91$ & 1.04 & 12.6\\
15 & $-127,-22$ & EPRB4,WS32 & $-166.$ & 10. & 31. & 61. & 45. & 9.57 & $-67$ & 27.5 & 45.6\\
16 & $-138,-57$ &  & $-165.$ & 5.5 & 13. & 7.3 & 7.0 & 1.25 & $119$ & 0.617 & 1.64\\
17 & $153,-58$ & EPRB46,WS16,17,18 & $-182.$ & 8.2 & 3.8 & 4.8 & 12. & 14.0 & $-4$ & 0.624 & 16.2\\
18 & $249,145$ &  & $-224.$ & 6.0 & --- & --- & 3.4 & 4.70 & $-176$ & --- & 3.08\\
19 & $-4,-340$ & EPRB20 & $-134.$ & 4.8 & 49. & 22.\tablenotemark{d} & 23. & 0.685 & $52$ & 4.99 & 0.0282\\
20 & $233,-26$ & EPRB7 & $-195.$ & 6.0 & 14. & 9.4 & 15. & 8.74 & $-17$ & 4.96 & 10.5\\
21 & $-99,-386$ & EPRB22 & $-119.$ & 5.8 & 42. & 27.\tablenotemark{d} & 12. & 4.40 & $137$ & 24.2 & 20.4\\
22 & $291,144$ & EPRB58 & $-210.$ & 4.6 & 10. & 4.0 & 5.5 & 6.75 & $93$ & 2.04 & 10.4\\
23 & $290,-28$ & EPRB72 & $-196.$ & 5.1 & --- & --- & 6.2 & 7.77 & $-123$ & --- & 11.2\\
24 & $211,490$ & EPRB10 & $-258.$ & 4.2 & 14. & 4.7 & 6.7 & 4.05 & $-5$ & 2.47 & 4.75\\
25 & $201,496$ & EPRB10 & $-256.$ & 6.7 & 18. & 15. & 15. & 7.24 & $-109$ & 7.26 & 13.9\\
26 & $-136,358$ & EPRB15 & $-213.$ & 7.2 & 7.8 & 7.7 & 8.5 & 11.9 & $-140$ & 2.24 & 12.0\\
27 & $-129,378$ & EPRB15 & $-216.$ & 6.0 & 16. & 11. & 5.5 & 1.54 & $112$ & 1.16 & 4.96\\
28 & $256,-199$ & EPRB6 & $-160.$ & 6.5 & 32. & 25. & 27. & 3.49 & $-24$ & 10.7 & 6.92\\
29 & $330,-19$ & EPRB21 & $-191.$ & 7.4 & 38. & 40. & 21. & 8.91 & $-170$ & 40.0 & 47.0\\
30 & $-129,395$ & EPRB15 & $-220.$ & 7.3 & 26. & 26. & 16. & 9.44 & $-157$ & 19.3 & 33.9\\
31 & $103,547$ & EPRB2 & $-243.$ & 6.5 & 29. & 23. & 32. & 3.58 & $-86$ & 9.06 & 17.6\\
32 & $132,559$ &  & $-244.$ & 5.2 & --- & --- & 5.8 & 4.76 & $63$ & --- & 4.31\\
33 & $268,526$ & EPRB74 & $-246.$ & 8.5 & 28. & 38.\tablenotemark{d} & 8.0 & 5.35 & $-55$ & 12.4 & 18.7\\
34 & $211,562$ &  & $-253.$ & 7.4 & 8.8 & 9.2 & 7.5 & 13.8 & $69$ & 3.29 & 16.7\\
35 & $90,562$ &  & $-246.$ & 5.4 & 7.9 & 4.3 & 6.2 & 9.35 & $-135$ & 1.77 & 19.1\\
36 & $115,579$ &  & $-253.$ & 6.7 & 47. & 41.\tablenotemark{d} & 24. & 3.96 & $-54$ & 26.9 & 45.8\\
37 & $238,570$ & EPRB1 & $-248.$ & 9.6 & 38. & 67. & 78. & 7.37 & $24$ & 33.3 & 27.6\\
38 & $218,625$ & EPRB108 & $-249.$ & 6.8 & 16. & 14. & 7.1 & 9.67 & $116$ & 8.03 & 14.6\\
39 & $295,-352$ & EPRB5 & $-157.$ & 9.0 & 21. & 32. & 32. & 7.47 & $-110$ & 9.61 & 29.3\\
40 & $548,413$ &  & $-248.$ & 6.2 & 13. & 9.3 & 9.6 & 7.93 & $-144$ & 3.85 & 12.8\\
41 & $552,431$ & EPRB8 & $-242.$ & 8.7 & 19. & 27. & 28. & 6.97 & $175$ & 7.66 & 11.4\\
42 & $564,403$ & EPRB9 & $-222.$ & 9.3 & 23. & 38. & 30. & 8.22 & $-130$ & 13.7 & 28.3\\
43 & $-362,-821$ &  & $-74.0$ & 6.3 & --- & --- & 4.6 & 9.95 & $-104$ & --- & 7.18\\
44 & $618,48$ & EPRB11 & $-202.$ & 6.4 & 18. & 14. & 6.7 & 10.8 & $-30$ & 10.7 & 23.4\\
45 & $627,65$ & EPRB11 & $-204.$ & 9.2 & 23. & 38. & 25. & 8.56 & $32$ & 14.4 & 19.2\\
\tableline
\end{tabular}
\begin{minipage}[h]{6.5in}
\tablenotetext{a}{Right ascension and declination are given in seconds
of arc relative to the galactic center at $\alpha_{2000}=
1^{\mbox{\tiny h}} 33^{\mbox{\tiny m}}
50\fs 8$ and $\delta_{2000}=30^{\circ} 39\arcmin 36\farcs 7$.}
\tablenotetext{b}{Names are those found in \citet{ws90}, denoted WSXX
where XX is the cloud number in Table 2 of their paper and those found in
Paper I, denoted EPRBXX where XX is the number in that paper.  Clouds
referred to twice are clouds resolved into two separate clouds by
higher angular resolution.}
\tablenotetext{c}{Properties are derived using methods described in
text.  Approximate errors are $\delta r = 5$ pc, $\delta
M_{\mbox{VT}}=$20\%, $\delta M_{\mbox{CO}}$=40\%, $\delta |\nabla v|$=0.02 km
s$^{-1}$ pc$^{-1}$., $\delta\phi$=10$^{\circ}$, $\delta j_1$=0.03 pc km
s$^{-1}$.}
\tablenotetext{d}{Filling fraction ($f$, Equation \ref{fillfrac}) is less
than 0.5, so the virial mass estimate is unreliable.}
\end{minipage}
}
\end{center}
\end{table*}

\subsection{Cloud Masses}
\label{mass_sec}
We measure the GMC masses using two methods: (1) assuming a constant
CO-to-H$_2$ conversion factor and (2) applying the virial theorem.
The CO-to-H$_2$ conversion factor \citep[$X$,][]{bloemen} converts
between integrated CO line intensity ($L_{\mathrm{CO}}$) and molecular
hydrogen column density ($N(\mathrm{H}_2)$).  We use a conversion
factor of
\[ X= 2 \times 10^{20} 
\frac{ N(\mathrm{H}_2)/\mathrm{cm}^{-2}} {L_{\mathrm{CO}}/\mathrm{K\
km\ s}^{-1}}\] 
\citep{sm96,dht01}.  In converting the column density
to a cloud mass, we include the mass of helium with a number fraction
of 9\% relative to hydrogen nuclei.  This gives the luminous mass of
the cloud:
\begin{equation}
\frac{M_{\mathrm{CO}}}{M_{\odot}}= 4.3 \left(\frac{X}{2\times
10^{20}}\right)\left(\frac{L_{\mathrm{CO}}}{\mathrm{K\ km\
s}^{-1}}\right)\left(\frac{A_{\mathrm{cl}}}{\mathrm{pc}^2}\right),
\end{equation}
where $A_{\mathrm{cl}}$ is the projected surface area of the molecular
cloud.  Our observations typically have $\sigma_{rms} = 1.0 \mathrm{\
K\ km\ s}^{-1}$ (4.3 $M_{\odot}\mbox{ pc}^{-2}$). Errors in measuring
masses using the $X$ factor arise not only from the uncertainty in the
$X$ factor but also from the flux calibration and the noise in the
data.  Determining the true masses of entire GMCs is further
complicated by the presence of low surface brightness emission around
the central clouds.

For the 36 resolved clouds, we can also compute the mass using the
virial theorem.  The virial mass requires a measurement of the
velocity width of the GMC and the cloud size.  We used the
intensity-weighted second moments of the position and the velocity to
determine the size and linewidth respectively, following \citet[][
hereafter S87]{srby87}.  The radius of the cloud $r$ is defined as
\begin{equation}
r=1.36 \sqrt{\sigma_{\alpha}^2+\sigma_{\delta}^2-2\sigma_{bm}^2}
\end{equation}
where $\sigma_{\alpha}$ and $\sigma_{\delta}$ are the dispersions in
right ascension and declination converted to projected distance and
$\sigma_{bm}$ is the beam width.  The coefficient scales the result to
match the definitions of cloud radius given in S87 to facilitate
comparison with Milky Way GMCs (\S\ref{comparison}).  The scaling is
chosen so that the radius of a circular cloud would be the same using
either measurement method.  Our size measure is independent of the
orientation of the coordinate system in which the moments are
calculated.  With the noise level of our data, this radius measurement
works best for cloud radii larger than 10 pc, the projected beam
radius.  With this definition of $r$, the virial mass of the cloud is
then:
\begin{equation}
M_{\mathrm{VT}}=1427 \left(\frac{\sigma_v^2}{\mbox{km s}^{-1}}\right) 
\left(\frac{\sqrt{\sigma_{\alpha}^2+\sigma_{\delta}^2-
2\sigma_{bm}^2}}{\mbox{pc}}\right).
\end{equation}
This virial mass estimate implicitly assumes a density profile of
$\rho \propto r^{-1}$.  These methods reproduce the virial masses
found by \citet{ws90} to within 40\% for the four clouds that are
well-resolved in both samples.

\begin{figure}
\begin{center}
\plotone{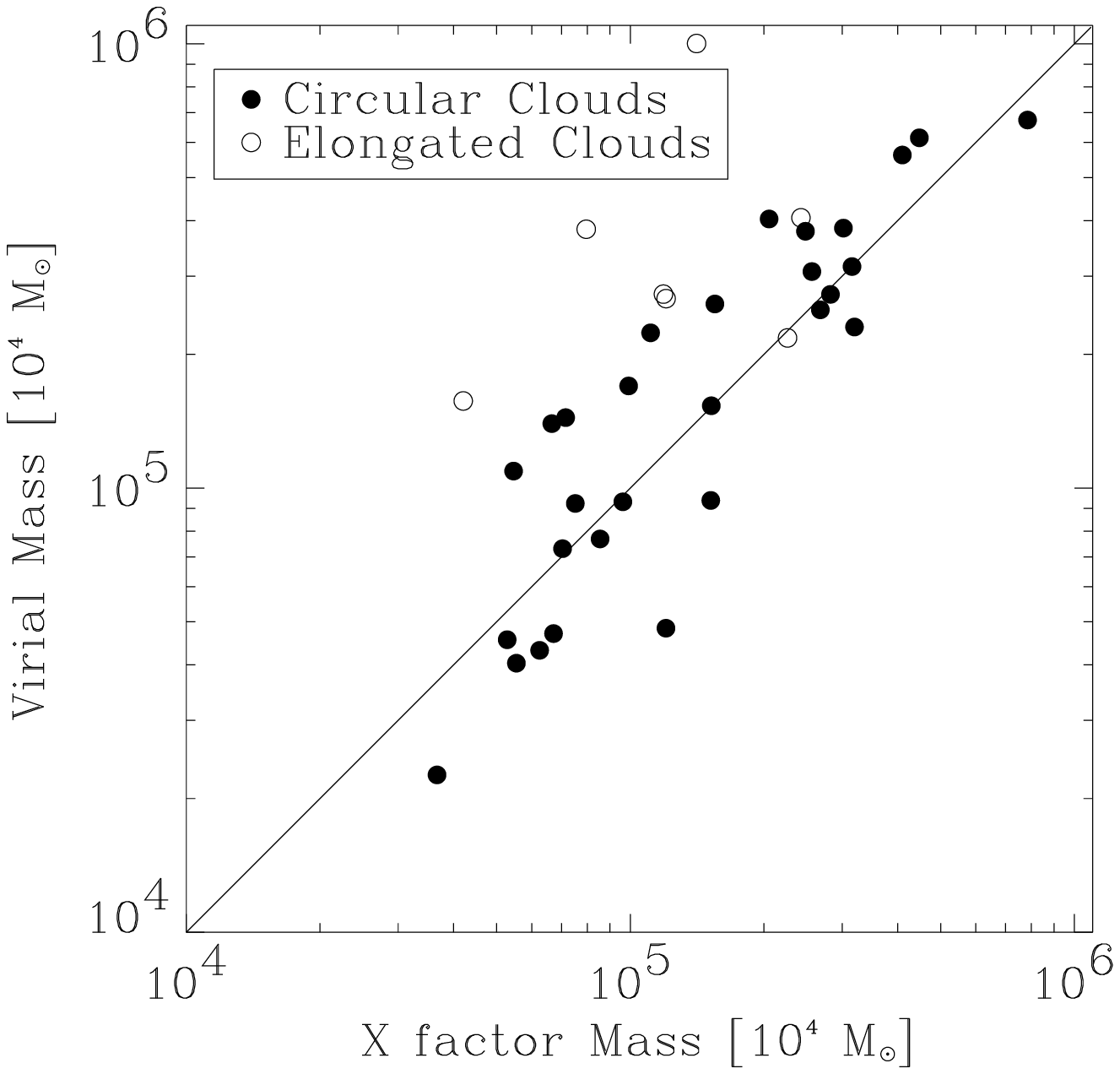} \figcaption{Comparison of virial and luminous
masses.  Elongated clouds are those with a spatial filling fraction
less than 0.5.  Errors for $M_{\mathrm{VT}}$ are $\pm 30\%$ and for
$M_{\mathrm{CO}}$ are $\pm 40\%$. The solid line indicates the locus
where $M_{\mathrm{VT}}=M_{\mathrm{CO}}$.  Clouds with a spatial
filling fraction greater than 0.5 show good agreement between their
virial masses and the masses derived from the $X$ factor with $X =
2\times 10^{20} \mbox{ H}_2 \mbox{ cm}^{-2}/(\mbox{K} \mbox{ km}
\mbox{ s}^{-1})$.
\label{masscomp}}
\end{center}
\end{figure}

The virial mass estimate assumes a spherically symmetric cloud so that
the velocity dispersion in the radial direction corresponds to the
size scale in the plane of the sky.  Therefore, the most reliable
virial mass estimates come from clouds that appear round on the sky.
We define the spatial filling fraction $f$ as \begin{equation}
\label{fillfrac}
f\equiv\frac{4 A_{\mathrm{cl}}}{\pi \ell_{\mathrm{max}}^2}
\end{equation}
which compares the area actually occupied by the cloud to the area
that the cloud would occupy if it were circular with diameter equal to
$\ell_{\mathrm{max}}$, the cloud's largest linear dimension.  We
restrict analyses involving virial masses to the 29 ``round'' clouds,
defined as having $f > 0.5$, to minimize errors in the virial mass
estimates.  Clouds with small filling fractions appear as elongated
structures and suggesting that they may be decomposed into multiple,
dynamically distinct GMCs ({\it e.g.} Clouds 5, 10, 19, or 21 in
Figure \ref{maps}).  Figure \ref{masscomp} compares the virial and
luminous ($X$ factor) masses for the 36 resolved clouds in our study.
A linear fit between the virial and CO masses for the 29 round clouds
finds $M_{\mathrm{VT}}=(1.0 \pm 0.1) M_{\mathrm{CO}}-(0.1\pm
0.2)\times 10^4\ M_{\odot}$ with a reduced $\widetilde{\chi^2}$ value
of 0.9.  The slope of unity implies that our adopted value for the $X$
factor is consistent with the virial masses.

In \S \ref{fluxrec}, we found that our observations underestimated the
integrated intensity and linewidth of the CO emission by 50\% and 25\%
respectively.  Recovering more of the flux with higher sensitivity
observations would increase the luminous mass as well as the size and
linewidth used in the virial mass.  These increases tend to offset
each other so the ratio of the luminous mass to the virial mass
remains roughly constant.  To investigate how omitting 50\% of the
flux in a field affects our mass estimates, we modeled the clouds as
azimuthally symmetric brightness distributions with a power law
surface brightness ($L_{\mathrm{CO}}(r)\propto r^{\omega}$).  The
linewidth scales with the size following the relation found in \S
\ref{comparison}.  Using the above methods, we calculated
$M_{\mathrm{CO}}/M_{\mathrm{VT}}$ for the central 50\% of the emission
and for the entire cloud.  For all values of $\omega$ between $-2$ and
0, the ratio for the central 50\% is never more than 1.5 times higher
than the ratio for the entire cloud.  This estimate is an upper limit
since some of the emission is likely from dynamically distinct, low
mass clouds surrounding each GMC, implying the interferometer recovers
significantly more than half the emission from the central cloud.  We
conclude that there is a good correspondence between virial and
luminous masses derived from the interferometer data.

The values of masses, radii and linewidths, are comparable to those
found in the Milky Way (S87).  The empirical relations between these
quantities also match those found for the Milky Way molecular clouds
and this similarity is discussed in detail in \S\ref{comparison}.

\subsection{Velocity Gradients}
\label{velgrad}

The high resolution CO observations allow measurement of the velocity
gradient across the molecular clouds.  Many mapping studies of Milky
Way molecular clouds note such gradients and are summarized in
\citet{p99}.  The clouds in M33 also show significant velocity
gradients, with magnitudes comparable to Milky Way clouds.  To measure
the gradient, we derive the velocity centroid at each position in
the cloud using the intensity weighted first moment of the masked
spectrum.  Then, we fit a plane to the resulting velocity centroid
surface.  The coefficients of the fit define $\nabla v_{\mathrm{c}}$,
the gradient in this surface and $v_r$, the radial velocity of the
cloud, after \citet{gbmf93}.  The vector quantity $\nabla
v_{\mathrm{c}}$ can also be expressed as the gradient magnitude,
$|\nabla v_{\mathrm{c}}|$, and the position angle of the gradient
$\phi_{cloud}$.  The latter is measured east from north to facilitate
comparison with the orientation of the galaxy.  Errors in quantities
derived from the fit are propagated from the original data using the
covariance matrix.

\begin{figure}
\begin{center}
\plotone{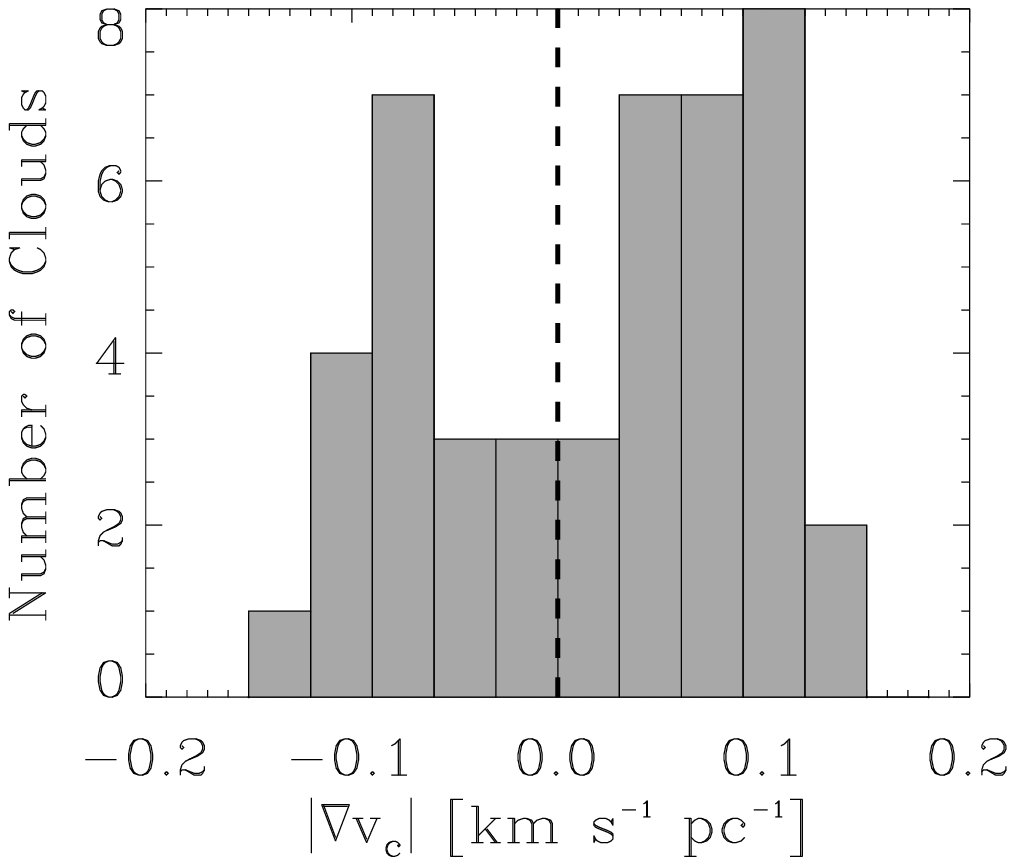} 
\figcaption{\label{gradhist} Histogram of gradient magnitude values
for clouds in M33.  Negative values are given to those clouds that
have a position angle differing from the galaxy by more than
$90^{\circ}$.  The gradient magnitudes are comparable to typical
values found in the Milky Way.  Moreover, the magnitudes of the
gradients are comparable among clouds independent of alignment with
the galaxy.}
\end{center}
\end{figure}

\begin{figure}
\begin{center}
\plotone{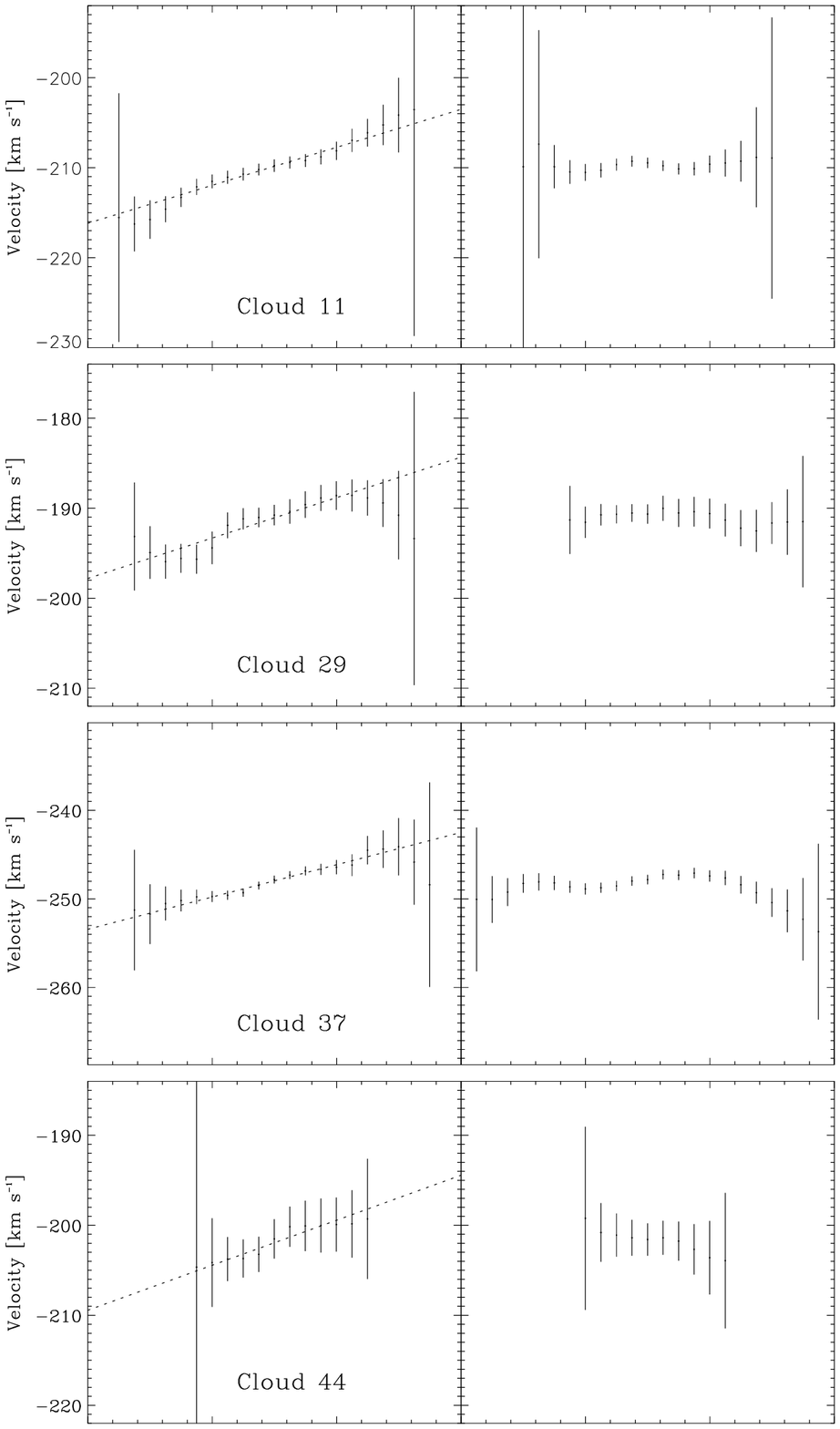} 
\vspace{0.5cm}
\figcaption{\label{gradfig} Examples
of four fitted gradient for molecular clouds observed with C-array.
The left-hand column contains plots of the velocity centroids denoted
by 1$\sigma$ error-bars from a position-velocity cut along the
direction of the gradient.  The dotted line is a linear fit to the
gradient.  The right hand column is identical to the left column,
except the position-velocity cut is made perpendicular to the
gradient.  The minimal curvature in the left column and zero slope in
the right indicate that planes are good fits to the velocity centroid
surfaces.}
\end{center}
\end{figure}

The results of the gradient fits are given in Table \ref{cloudprop}.
Planes are good fits to the velocity centroid surfaces with the median
value of reduced $\widetilde{\chi^2} = 1.4^{+2.2}_{-0.7}$ for the
measured gradients.  Fitting other functions (like paraboloids or
radial collapse profiles) to the centroid surface yield significantly
higher values of $\widetilde{\chi^2}$.  The magnitudes of the
gradients are displayed in the histogram in Figure \ref{gradhist}.
The magnitudes are comparable to the typical 0.1 km s$^{-1}$ pc$^{-1}$
for the clouds found in the Milky Way \citep{psp3,p99}.  In addition,
there is no significant relation between position angle and gradient
magnitude.  Figure \ref{gradfig} shows position-velocity cuts along
the derived gradients for four clouds in the data set.  Adjacent to
each plot is a position-velocity cut perpendicular to the derived
gradient.  This figure demonstrates that planes are good fits to the
data since the cut along the gradient shows a linear trend and the
perpendicular cuts appear constant.

\begin{figure}
\begin{center}
\plotone{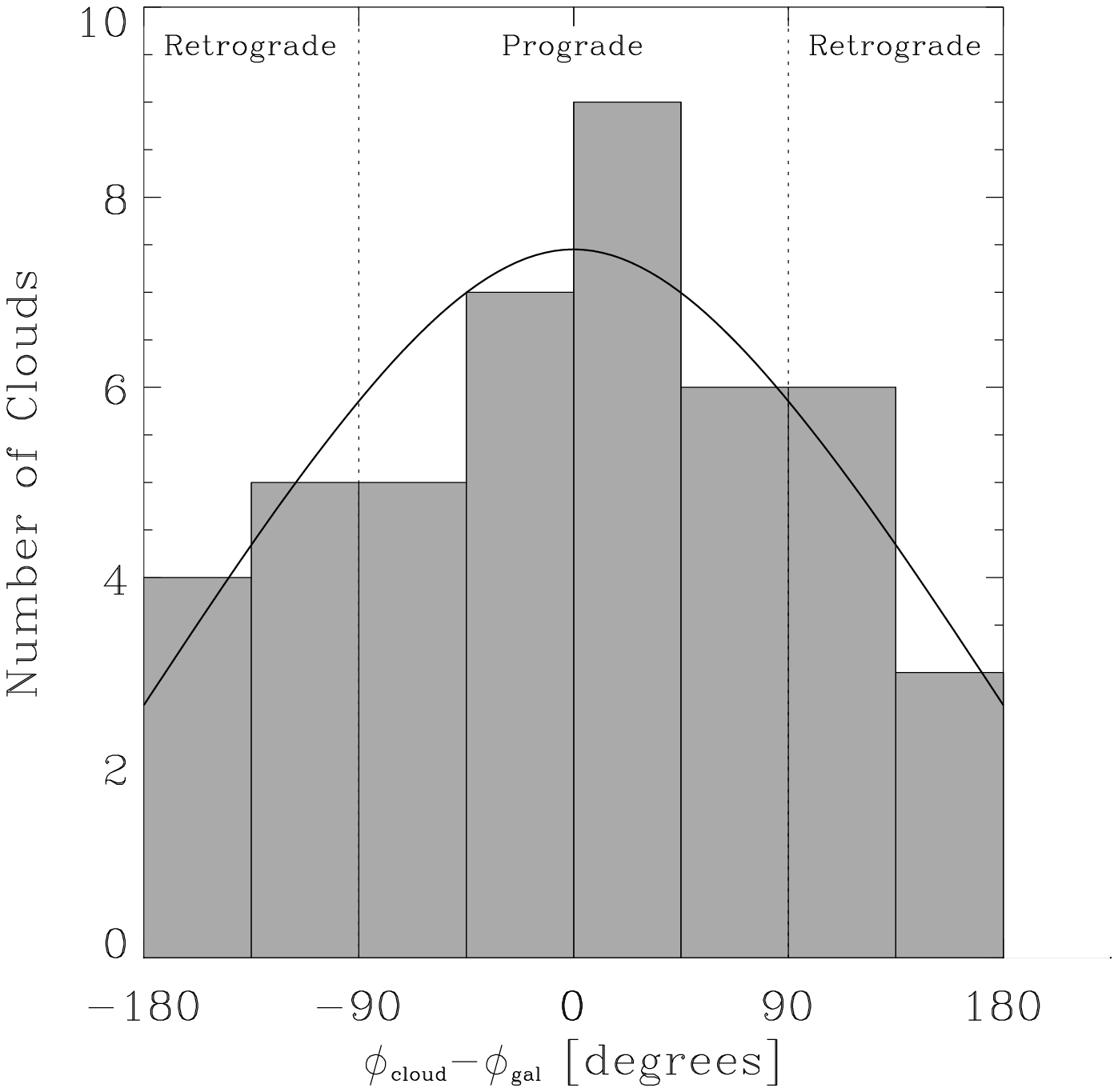} 
\figcaption{\label{pahist} Histogram of the position angles of the GMC
rotation with respect to that of the galaxy.  Clouds show slight
preferential alignment with the orientation of the galaxy. The line
represents the predicted distribution for a value of $\alpha=0.69$ in
Equation \ref{pdf}.  Rough separations between Prograde and Retrograde
clouds are shown, but the unknown inclinations of the clouds prevent
this separation from being completely determined.}
\end{center}
\end{figure}

Most authors assume that the velocity gradients are the signature of
cloud rotation \citep{psp3,p99}.  If the gradients are due to
rotation, then the magnitude of the velocity gradient measures the
angular velocity vector $\mathbf{\Omega}$ projected into the plane of
the sky $|\nabla v_{\mathrm{c}}|=|\mathbf{\Omega}| \sin i$.  The
position angle of the gradient is
$\phi_{cloud}=\phi_{\Omega}+90^{\circ}$.  The gradient therefore
provides much information about the angular velocity and rotation axis
of a cloud.  We find (1) the {\it linear} gradients observed are the
signature of {\it solid-body} rotation.  (2) The rotation energy for these
clouds is $\lesssim 5\%$ of the cloud binding energy.  (3) The
position angles of the rotation axes show marginal alignment with the
axis of the galaxy, as is seen in Figure \ref{pahist}.  (4) There are
many clouds with position angles separated by more than $90^{\circ}$
from that of the galaxy.  (5) If the clouds are rotating, roughly 40\%
of the clouds are retrograde rotators (\S\ref{pacorrs}). (6) For the
median gradient magnitude of $0.05$ km s$^{-1}$ pc$^{-1}$, the
rotation period of a molecular cloud would be 125 Myr, significantly
longer than the assumed molecular cloud lifetimes of 10 to 30 Myr
\citep[Paper I,][]{psp3}.

We measure the angular momentum of the molecular clouds from the
gradient fits and the cloud radii.  For a solid body rotator with a
power law surface mass density, the specific angular
momentum\footnote{Through the remainder of the paper, we will refer to
the specific angular momentum as the angular momentum.  Any case where
the true angular momentum, $J=jM$, is intended will be specifically
noted.}  $j = J/M = \beta |\nabla v_c| r^2$ where $\beta$ is a
constant set by the mass profile and $r$ is the radius of the cloud.
For a constant surface density cloud, $\beta=0.4$, ranging from 0.33
to 0.5 for different density profiles \citep{p99}.  This analysis
adopts $\beta=0.4\pm 0.1$.  The derived angular momentum is listed as
$j_1$ in Table \ref{cloudprop}.  The alignment of the rotation axis
with respect to the observer affects the magnitude of the observed
gradient.  The value $j_1 = j \sin i$ where $j$ is the true angular
momentum of the cloud and $i$ is the inclination to the line of sight.
Without knowing the inclination, $j_1$ will underestimate the true angular
momentum.  To avoid assuming solid-body rotation, we calculated a
second measurement of the angular momentum
\begin{eqnarray}
j_2 & = & \sum_{i,j,k} T_A(x_i,y_j,v_k) 
\left[(x_i-x_0)^2+(y_j-y_0)^2\right]^{1/2}\cdot \nonumber\\
& & (v_k-v_0) \sin \vartheta_{i,j} \cdot \left[\sum_{i,j,k} T_A(x_i,y_j,v_k)\right]^{-1}
\end{eqnarray}
where $x_0,y_0,$ and $v_0$, are the centers of the cloud averaged over
position and velocity space respectively and $\vartheta_{i,j}$ is the angle
between the point $(x_i,y_j)$ in the cloud and the gradient judged from
the center of the cloud.  This value can be calculated even when the
cloud is too small to be deconvolved accurately.  Even with this
measurement, a cloud can be oriented with its angular velocity vector
along the line of sight thereby obscuring any signature of rotation.
The effects of inclination must ultimately be dealt with statistically
under the assumption that the clouds are not preferentially oriented
towards the observer.

\citet[][BB00]{bb00} show that turbulent velocity fields can also
produce linear gradients in the observational domain.  The required
turbulent power spectrum, $P(k) \propto k^{n}$, has a value of $n$
between $-3$ and $-4$, so that the largest size-scales have the most
power, appearing as gradients.  Using their scaling for the magnitude
of the velocity gradient as a function of size gives gradients on
order 0.08 km s$^{-1}$ pc$^{-1}$ for the median cloud radius of 20 pc,
comparable to the observed values.  If the gradients are due to
turbulence and not rotation, our measurements change only in that the
angular momentum derived from the data will {\it overestimate} the
true value of the angular momentum by a factor of 2 to 3 (BB00).  The
orientation of the net angular momentum vector is still represented by
the position angle of the gradient.

\section{Discussion}
\label{disc}

\subsection{Angular Momentum Constraints on GMC Formation}

The measured angular momenta are significantly less than expected from
simple theories of cloud formation.  In the absence of external
forces, the angular momentum of a molecular cloud should be equal to
the angular momentum of the gas from which it formed.  In a shearing
galactic disk, the initial angular momentum for a forming cloud is
(see Appendix \ref{angder}):
\begin{equation}
\label{jgal}
j_{gal} = \eta \left.\frac{1}{R_c}\frac{d}{dR}(RV)\right|_{R=R_c}
\Delta R^2.
\end{equation}
In this equation, $R$ is the distance from the center of the galaxy,
$R_c$ is galactocentric radius where the cloud is found, and $V(R)$ is
the galactic rotation curve \citep[from][]{cs97}.  The radial
accumulation length $\Delta R$ is half the radial extent of the parent
atomic gas.  The radial mass distribution of the collapsing gas
determines the parameter $\eta$.

In the simplest case, the radial accumulation length is set by
requiring the mass of atomic hydrogen in a cylinder stretching to
infinity in the $z$ direction and of radius $\Delta R$ to equal the
mass of the molecular cloud: $\Sigma_{\mbox{\tiny \sc Hi}}\cdot \pi
\Delta R^2 = M_{\mathrm{GMC}}$. We adopt $\eta=1/4$ in Equation
\ref{jgal} calculated for the collapse of a cylinder.  Using the
measured masses of molecular clouds and the local value of
$\Sigma_{\mbox{\tiny \sc Hi}}$, we find that {\it in all but 3 of the
36 resolved molecular clouds, the predicted angular momentum is higher
than the observed value, on average by a factor of 5}.  We calculated
the values of $\Sigma_{\mbox{\tiny \sc Hi}}$ using the maps of
\citet{deul} after recalibrating this map to match the surface density
of \citet{cs97}.  If the velocity gradients are due to turbulence
instead of rotation, BB00 argue that the observed angular momentum
should be scaled down by a factor of 2 to 3, which only increases the
discrepancy between the predicted angular momentum and the observed
value.  Statistically, projection effects can account for only 20\% of
the discrepancy.

\begin{figure}
\begin{center}
\plotone{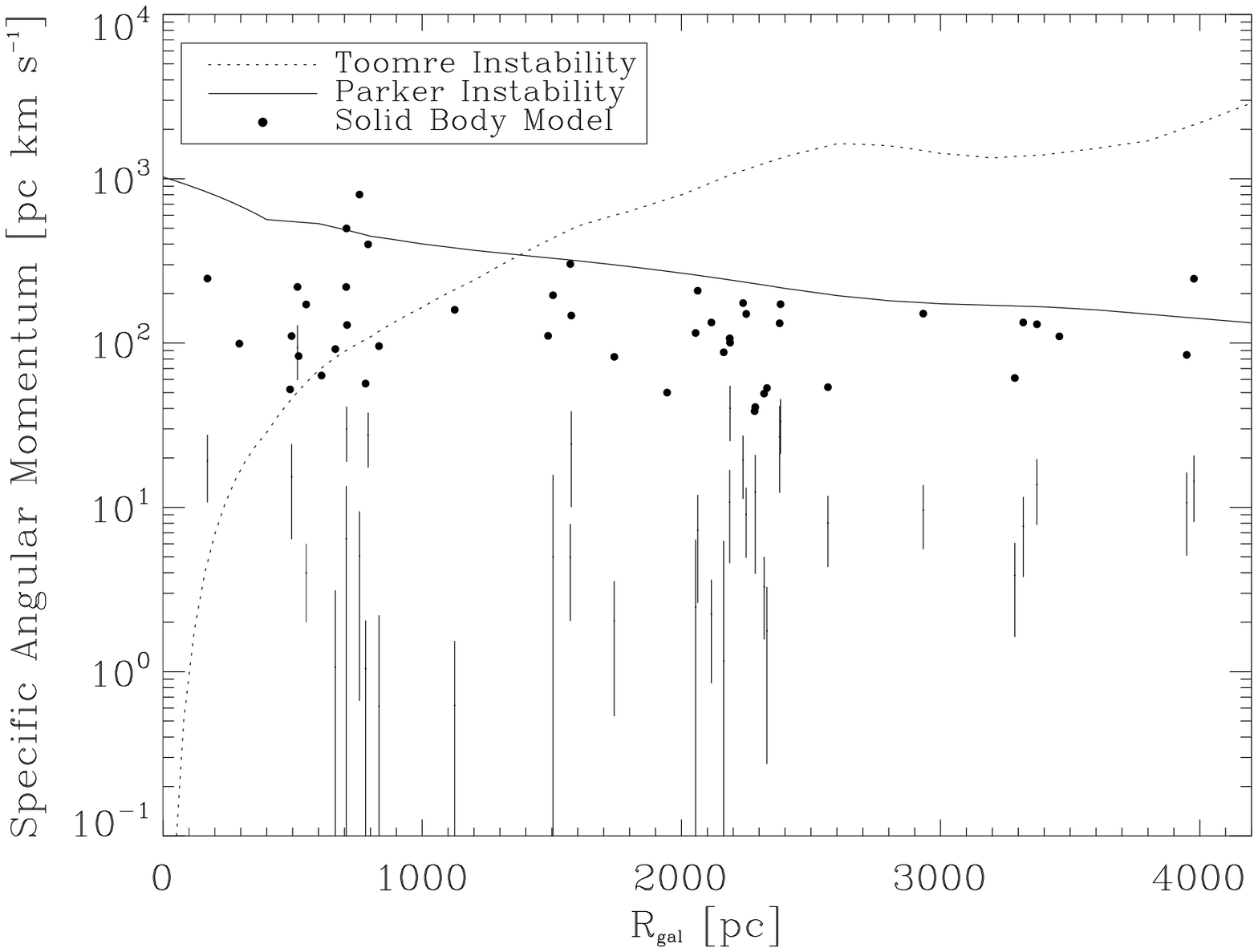} \figcaption{\label{spangcomp} Specific
angular momenta ($j_1$) imparted by different collapse mechanisms as a
function of galactic radius.  The measurements are represented by $\pm
1\sigma$ error bars.  The two curves are the imparted angular momenta
as a function of $R_{gal}$ for the Toomre and Parker Instabilities.
The filled points are the predicted values of the specific angular
momentum for the solid body rotator discussed in the text.  Nearly all
measurements of the specific angular momentum lie well below the
predictions for all models.}
\end{center}
\end{figure}

We refine this simple model as follows: (1) by assuming that the gas
far from the galactic plane does not contribute to the forming
molecular cloud and (2) by including some estimate of projection
effects.  Instead of the progenitor material extending infinitely far
from the galactic plane, we assume the gas is accumulated over a
cylinder with its height equal to its diameter, centered on the
location of the GMC.  The three-dimensional distribution of atomic
hydrogen is unknown, but we assume a scale height of $H=$100 pc
throughout the galaxy with $\rho(z)\propto \mbox{sech}^2(z/2H)$.
Changing this value by a factor of 2 does not alter the results
significantly.  Again, the dimensions of the cylinder are determined
by requiring its volume to contain a mass of atomic gas equal to that
of the molecular cloud.  Using the radius of the cylinder as the
radial accumulation length $\Delta R$, the predicted angular momentum
is determined from Equation \ref{jgal}. We also assume the clouds are
solid body rotators with random orientations and scale the predicted
angular momentum down by a factor of $\langle\sin i\rangle =\pi/4$.
If, instead, we assume the clouds are strictly
aligned with the galaxy, the correction only differs from $\pi/4$ by
3\%.  We compare this model to the measurements in Figure
\ref{spangcomp}.  The measured angular momenta are plotted as $\pm
1\sigma$ error bars as a function of galactic radius.  Figure \ref{spangcomp}
shows that the observations are always less than the predictions of
this Solid Body Model (filled circles).  The discrepancy is more than
an order of magnitude for 20 of the clouds.

\subsubsection{The Toomre and Parker Instabilities}

Recent work on star formation at galactic scales suggests that star
formation preferentially occurs where the gas disk is Toomre unstable
\citep{mk01}.  We calculate the radial accumulation length if the
clouds actually formed from the instability.  The \citet{toomre}
instability is a gravitational instability in a shearing, thin fluid
disk that bunches the gas into concentric rings, provided
the disk is unstable according to the criterion
\begin{equation}
\label{toom_eq}
\frac{\kappa \sigma_v}{\pi G \Sigma(R)} < 1
\end{equation}
where $\kappa$ is the local epicyclic frequency, $\sigma_v$ is the
(three dimensional) velocity dispersion of the fluid, and $\Sigma(R)$
is the local surface mass density of the fluid under consideration.
The radial accumulation length is set by the original radial extent of
the contracting rings:
\begin{equation}
\Delta R_{T}=\frac{2\pi^2 G
\Sigma(R)}{\kappa^2}=\frac{\lambda_T}{2},
\end{equation}
where $\lambda_T$ is the most unstable scale for Toomre collapse.  The
rotation curve determines the epicyclic frequency for the galaxy
\[ \kappa^2 = 2 \left( \frac{V^2}{R^2}+\frac{V}{R}\frac{dV}{dR}\right)\]
We assume that there is sufficient mass in the annulus to form
molecular clouds which is true over most of the galaxy.

Figure \ref{spangcomp} also shows the predicted values of the cloud
angular momentum for GMCs formed using the Toomre instability.  These
values only agree with the observed angular momenta for a few clouds
in the inner 500 pc of the galaxy, {\it where the disk is most
stable} according to the criterion in Equation \ref{toom_eq}.  Since
all but two of the clouds disagree with the predictions of the Toomre
instability by a large margin, it seems unlikely that
the clouds form by the action of this instability alone.  Moreover,
the Toomre instability predicts that cloud angular momentum should
increase with galactocentric radius, which is not seen in Figure
\ref{spangcomp}. 

The Parker instability is another potential avenue of molecular cloud
formation \citep{mouschovias,eg82,hanawa}.  The Parker instability is a
magneto-hydrodynamic (MHD) effect in gas coupled to a buoyant magnetic
field.  As the instability evolves, a section of the field bulges out
of the plane of the galaxy, causing the atomic gas to flow down the
gravitational potential well along the edge of the bulge and collect
into molecular clouds. The characteristic length scale of the
instability is $\lambda_P \approx \pi H$, where $H$ is the scale
height \citep{gs93}.  The values for the predicted angular momentum
are shown in Figure \ref{spangcomp}, for a 100 pc scale height.  These
values have been corrected for the pitch angle of the magnetic field
\citep[\ensuremath{\theta=60^{\circ}}, ][]{beck00} since the collapse
occurs along the magnetic field lines, so that $\Delta R_P=(\lambda_P
\sin \theta)/2$.  The Parker instability also gives a large discrepancy
between the observed and predicted values.  Matching the observed
values would require a gas scale height of $\sim 20$ pc, which would
imply an unrealistically large stellar mass density.

\subsubsection{The Angular Momentum Problem}
\label{mhdsec}

We conclude {\it none of the simple theories produce the observed
values of angular momentum in the molecular clouds}.  Figure
\ref{spangcomp} compares the theoretical and observed angular momentum
for the resolved molecular clouds.  Even using the moment based
measurement of angular momentum ($j_2$), we find an average
discrepancy of more than a factor of 5 between observation and theory.
This discrepancy shows that nearly all clouds suffer an angular
momentum problem independent of assuming solid-body rotation.  The
failure of the Solid Body Model is particularly troubling since this
model gives the minimum accumulation length required to gather the
mass of the GMC from the atomic hydrogen.  This constraint will limit
even the most sophisticated models of cloud formation.

There are at least two possible solutions to the angular momentum
problem.  One is to gather material for molecular clouds from similar
galactocentric radii, thereby minimizing the shear in the atomic gas,
as may happen in the swing amplification instability \citep{ko01}.  If
clouds have small radial accumulation lengths, they must have long
azimuthal accumulation lengths.  Agreement with the observed angular
momenta requires $\Delta R < 30$ pc.  To form a cloud of $10^5\
M_{\odot}$ out of atomic hydrogen with a typical surface density of
$7\ M_{\odot}$ pc$^{-2}$ requires azimuthal accumulation lengths of
$\sim$500 pc.  With the maximum non-circular velocities in the atomic
gas $\lesssim$ 10 km s$^{-1}$ \citep{deul}, this implies a limit on
the timescale of cloud formation of $\tau_{form} \gtrsim 50$ Myr,
larger than the 10---20 Myr cloud lifetimes implied in Paper I.  This
timescale is a significant fraction of the time between the spiral arm
crossings for most of the galaxy away from the co-rotation radius
\citep{m33_newton}.  Consequently, such a long time scale for cloud
formation has difficulties explaining preferentially finding clouds in
spiral arms of galaxies.

A second solution is to brake the material with external forces such
as magnetic fields.  Magnetohydrodynamic effects seem to be critical
for shedding angular momentum in the star formation process and may
also act in the early stages of cloud formation.  If the magnetic
field is so dynamically significant as to produce an instability, it
may also provide enough tension to brake the clouds during their
formation.  \citet{beck00} measures the mean field strength in M33 as
$6\pm 2 \mu \mathrm{G}$, which gives an Alfv\'en speed of $\sim$6 km
s$^{-1}$ (assuming a sech$^{2}$ density profile to calculate the
midplane volume mass density).  This value is comparable to the sound
speed in the warm neutral medium of most galaxies, suggesting that MHD
braking and cloud formation via instabilities occur on comparable
timescales.  The braking time is set by the time needed for an
Alfv\'en wave to sweep through a volume of diffuse gas with a moment
of inertia equal to that of the forming cloud \citep{braking_time}.
If the atomic gas has a constant density, then the Alfv\'en wave must
pass through a distance comparable to the accumulation length for the
molecular cloud.  Instabilities cause cloud formation on a time scale
comparable to the sound crossing time for the accumulation
length. Since the crossing times for sound and Alfv\`en waves are
comparable, we conclude that MHD braking is a viable option to slow
the rotation of forming clouds.  Because only 10\% of clouds show
angular momentum approaching the expected values from any of the
formation mechanisms (e.g. Figure \ref{spangcomp}), the MHD braking
must occur in the atomic gas or during a small fraction of the
molecular cloud lifetime.  We return to the implications of these
observations for cloud formation in \S\ref{finis}.

\subsubsection{Angular Momentum Magnitude Correlations}
\label{magcorr}
\begin{figure}
\begin{center}
\plotone{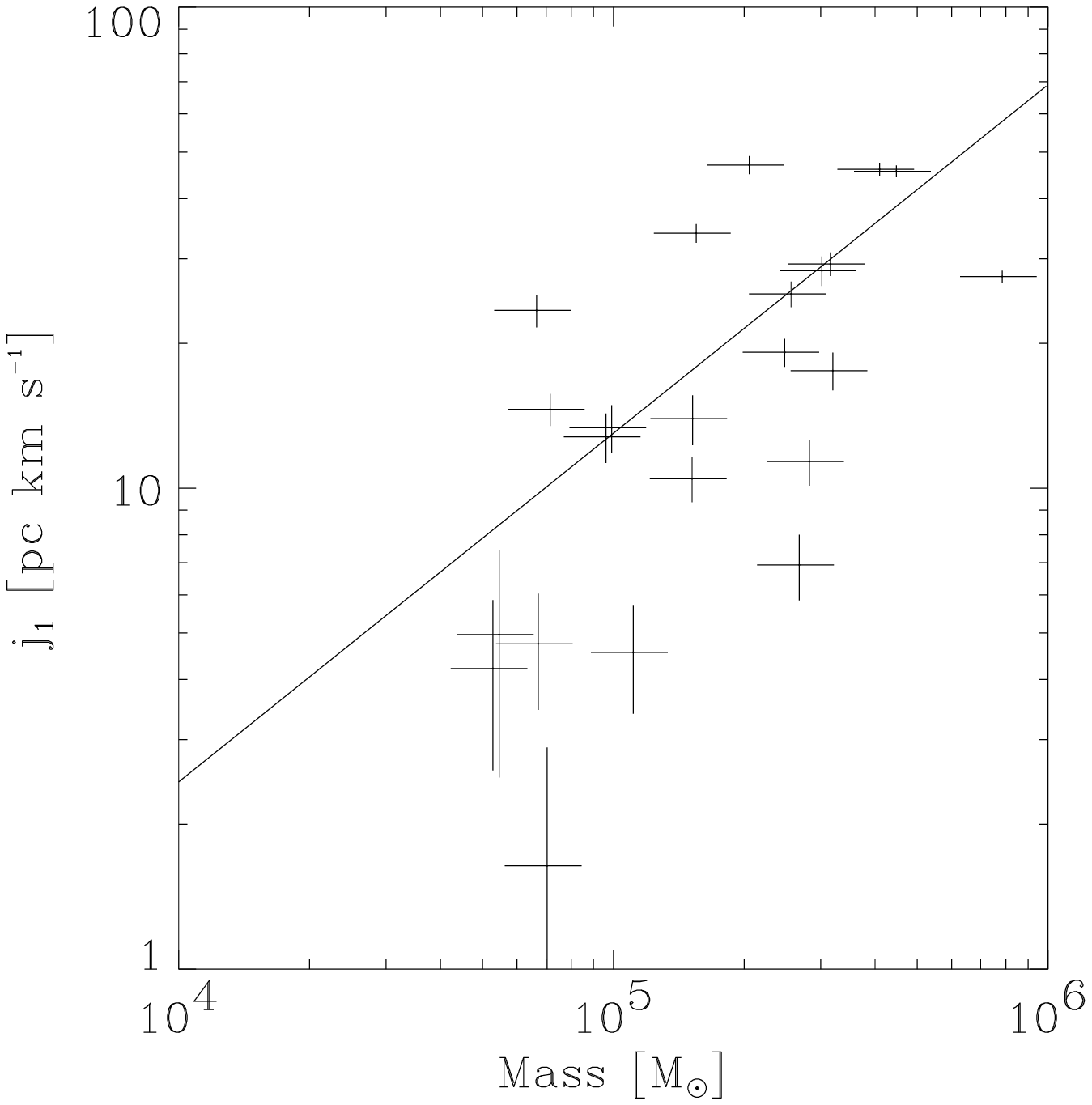} \figcaption{\label{mags} The correlation between the
magnitude of the specific angular momentum and the mass of the cloud.
Only clouds that are well resolved with low filling factors are
included in this fit.  The solid line is a least-squares, power-law
fit to the data with $j_1 \propto M^{0.6\pm 0.1}$ accounting for
errors in both directions.}
\end{center}
\end{figure}

The molecular clouds in this sample suggest a correlation between
specific angular momentum and mass.  Fitting a power-law relationship
between $j_1$ and $M_{\mathrm{CO}}$ finds $j(M)\propto M^{0.6 \pm
0.1}$, though there is significant scatter in the data
($\widetilde{\chi^2}=6.1$, Figure \ref{mags}).  A similar correlation
appears in the Milky Way data where \citet{p99} finds $j \propto
M^{0.7}$.  If the molecular clouds form by accumulation of atomic gas,
then the angular momentum should scale as the accumulation length
squared (Equation \ref{jgal}): $j \propto \Delta R^2$.  The mass of
the cloud is $M=\pi\Delta R^2\overline{\Sigma_{\mbox{\tiny \sc Hi}}}$,
so a correlation is expected with $j \propto
M/\overline{\Sigma_{\mbox{\tiny \sc Hi}}}$.  The variations in the
mean hydrogen surface density are small compared to the order of
magnitude spanned in mass (\S\ref{comparison}), so the angular
momentum is expected to scale with mass for any accumulation model.
That $j\propto M^{0.6}$ instead of $j\propto M$ may be due to the
action of the braking mechanism, though $j\propto M$ is not strongly
ruled out by the poor quality of the fit.  Such correlation between
$j$ and $M$ is also expected for a turbulent velocity field.  BB00
find that the velocity gradient scales with clouds size $r$ as
$|\nabla v| \propto r^{-0.5}$ so $j_1\propto |\nabla v| r^2 \propto
r^{1.5}$.  The clouds in our sample have a constant surface density
regardless of cloud size (\S\ref{comparison}).  Therefore, the mass of
a cloud is proportional to its area $M\propto r^2$ and $j\propto
M^{0.75}$, close to the observed trend.

There is no significant variation of specific angular momentum as a
function of galactic radius, contrary to the predictions of all the
accumulations mechanisms considered here (Figure \ref{spangcomp}).
This implies that the angular momentum of the clouds does not depend
on any of the properties that change across the galactic disk,
including galactic shear, surface mass density of stars and atomic
gas, interstellar pressure, interstellar radiation field, or
metallicity.

\subsubsection{Position Angle Correlations}
\label{pacorrs}
The position angle of the angular momentum vector can be measured from
the velocity gradients as discussed in \S \ref{velgrad}.  The data
show correlations between the cloud position angles and that of the
galaxy as well as among the individual GMCs.  If the angular momentum
is set entirely by the galactic shear, then the rotation axes of the
GMCs should be aligned with the galaxy.  The distribution of gradient
position angles is shown in Figure \ref{pahist}, including rough
divisions between prograde and retrograde rotators (assuming the
gradients represent rotation).  Because of the unknown inclination of
the clouds relative to the line of sight, we analyze the distribution
of position angles statistically.  We assume that the clouds are
distributed with orientations on the sphere set by the probability
distribution function (PDF):
\begin{equation} 
\mathbb{P}(\theta,\varphi)=\frac{1}{2\pi\alpha} \left[\frac{\cos 
\theta +1}{2}\right]^{1/\alpha-1} \frac{\sin \theta}{2}\label{pdf}
\end{equation} 
where $\theta$ is the angle between the angular momentum vector of the
cloud and that of the galaxy, $\alpha$ is a parameter that determines
the degree of alignment between these vectors, and $\varphi$ is the
azimuthal angle.  For $\alpha=1$, the orientation angles are randomly
distributed on the sphere so there are an equal number of prograde and
retrograde rotators. For $\alpha \approx 0$, the clouds are closely
aligned with the galaxy ($\alpha=0.2$ implies that 91\% of the clouds
are prograde rotators).  This PDF is motivated solely for ease of
simulation so that a single parameter $\alpha$ determines how well
aligned the clouds are with the galaxy.  By determining the value of
$\alpha$ that most closely represents the observed distribution of
position angles, we measure the degree of alignment between the clouds
and the galaxy.  We generated position angle distributions following
this PDF using Monte Carlo simulations and then used a one-sided K-S
test to measure the difference between the observed and simulated
distributions.  After testing a range of $\alpha$, we found the most
probable value is $\alpha=0.69$ (with K-S likelihood $P=0.99$),
implying 61\% of the clouds are prograde rotators.  The likelihood of
a random distribution on the sphere ($\alpha=1$) is significantly
lower ($P=0.60$) implying some alignment between the clouds and the
galaxy.  Nevertheless, the wide range of position angles indicates any
angular momentum imparted by galactic shear is randomized in some
fashion.

\begin{figure}
\begin{center}
\plotone{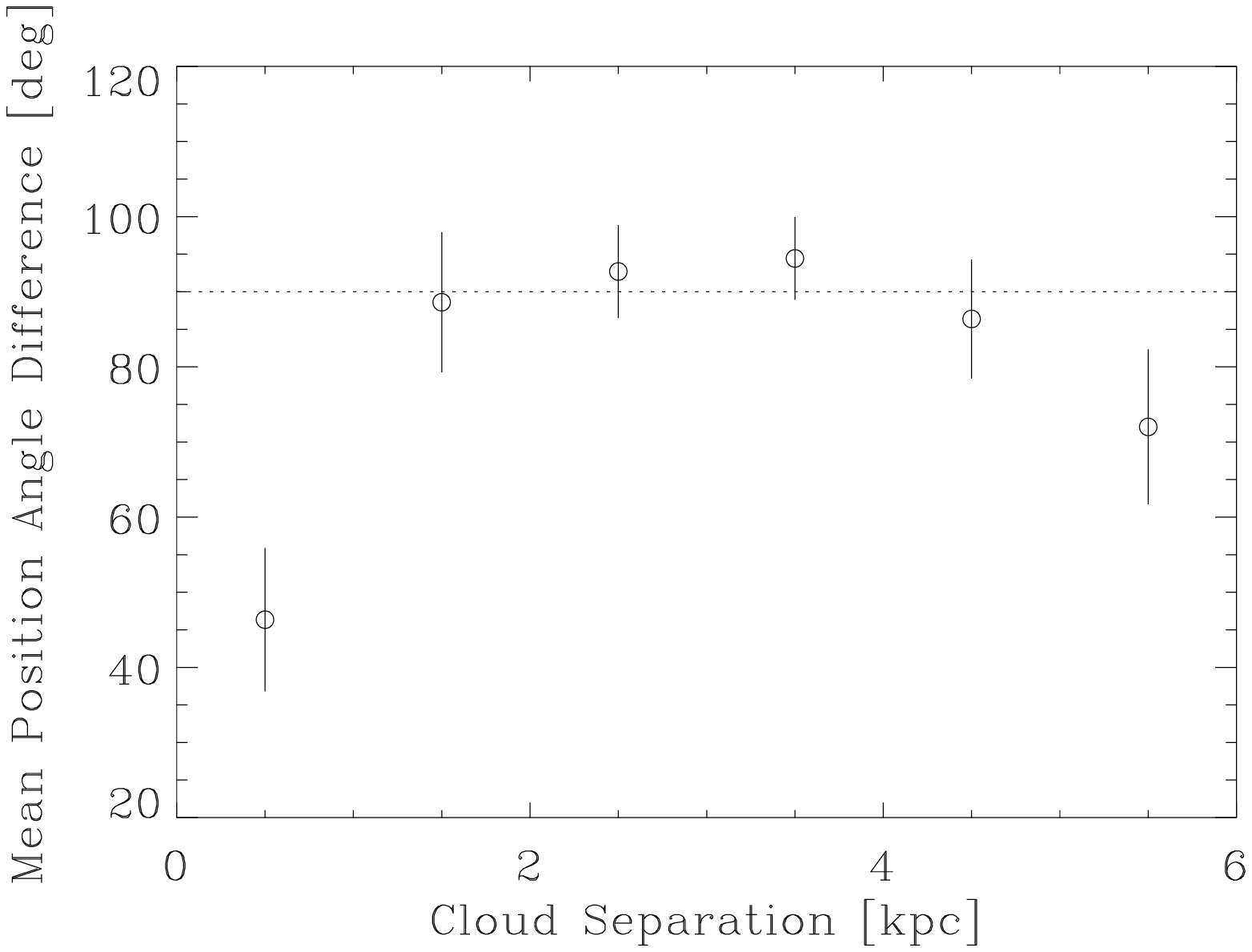} 
\figcaption{\label{pacorr} The correlation
between the position angle of molecular clouds and their separation
for clouds with $M > 8\times 10^4 M_{\odot}$.  There is a trend that
clouds at small distances tend to have their velocity gradients
aligned with each other. The mean in each bin is weighted according to
the uncertainty in the measurements and the error bars are the error in the
mean.  The first bin is the average of 21 pairs of position angles.}
\end{center}
\end{figure}

In addition to the cloud-galaxy correlation, nearby clouds show a
higher degree of correlation between their gradient position angles.
Figure \ref{pacorr} shows the difference in position angle between
pairs of clouds as a function of cloud separation.  This figure uses
clouds with $M > 8 \times 10^{4} M_{\odot}$ since these clouds have
the most significant gradients.  If the clouds are randomly oriented
with respect to each other, the value of this statistic should be
$90^{\circ}$, but clouds with separations smaller than 500 pc show a
significantly smaller value implying their angular momentum vectors
are aligned.  This alignment would be expected if these neighboring
clouds formed from the same parent cloud of \ion{H}{1}.  If clouds are
the product of large scale motions in the diffuse ISM, this 500 pc
length may represent the scale of the flows that give rise to the
clouds.

\subsection{Larson's Laws in M33}
\label{comparison}

The clouds in the M33 observations have similar radii ($\sim 20 \to
50$ pc), masses ($\sim 10^4 \to 10^6\ M_{\odot}$) and velocity
dispersion FWHMs (5 $\to$ 10 km s$^{-1}$) as those found in the Milky
Way by S87.  We tested these similarities in more detail and found the
GMCs in M33 were indistinguishable from those in the Milky Way.  We
emphasize comparison of our results with the work done by S87 because
both studies use equivalent methods for determining the properties of
the molecular clouds and the sensitivities are comparable: $\sim$0.7 K
km s$^{-1}$ for S87 vs. 0.6 $\to$ 1.3 K km s$^{-1}$ for this work.  We
only use the 23 clouds that are well-resolved ($r > 10$ pc) and round
($f \geq 0.5$) in these comparisons since the derived properties of
these clouds are the most reliable.

\begin{figure*}
\begin{center}
\plottwo{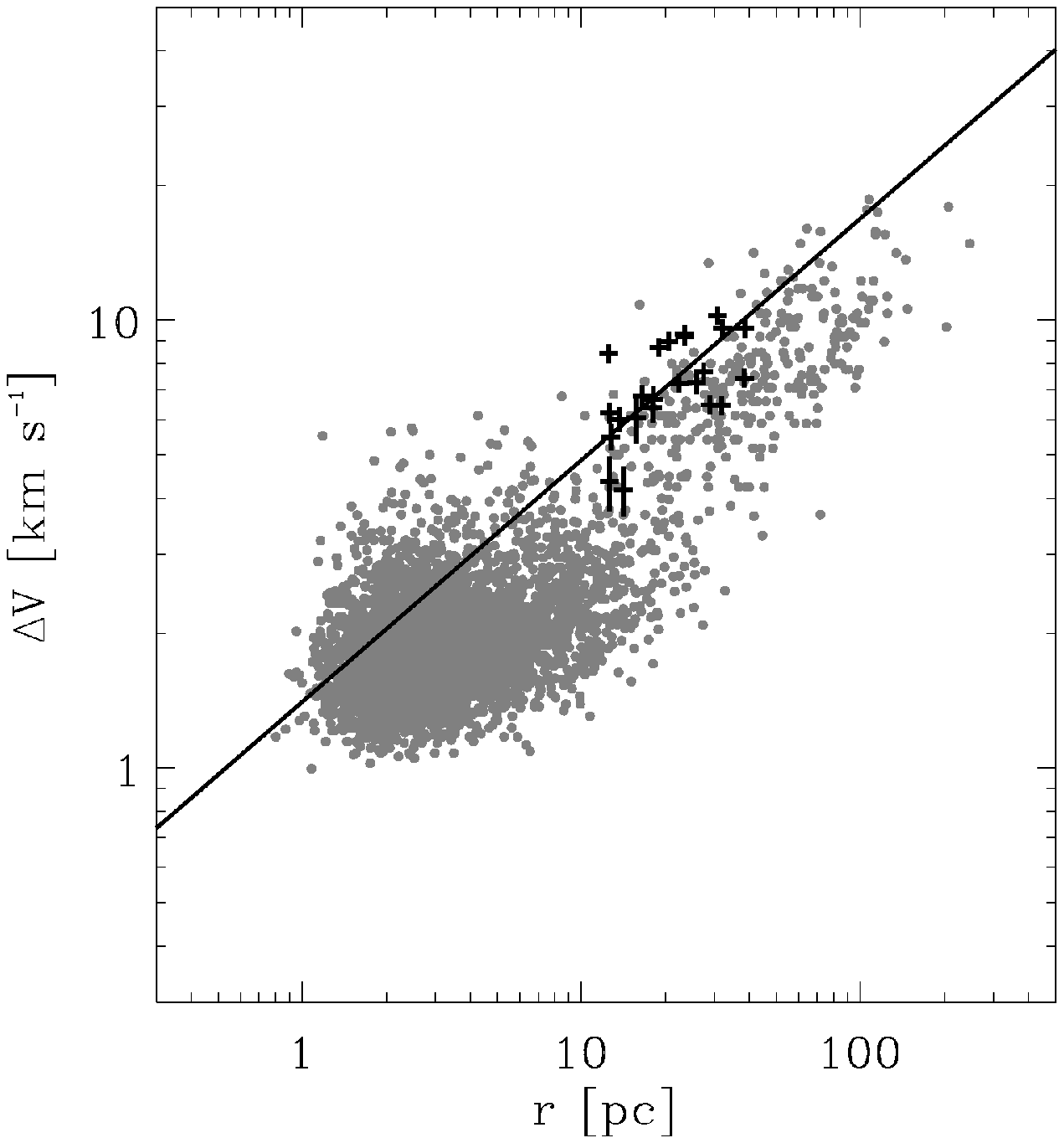}{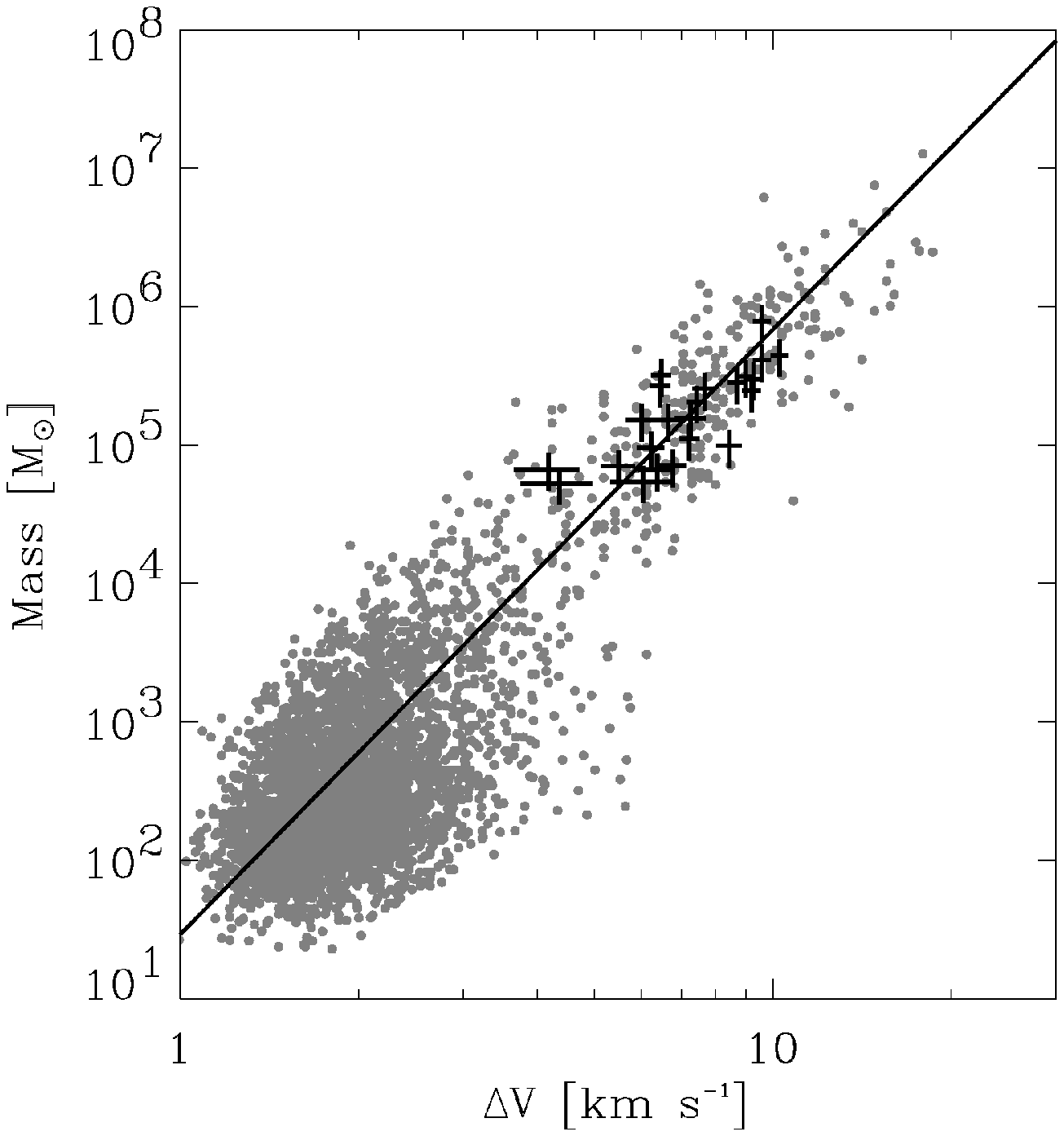} \figcaption{\label{larson} The
size-linewidth and mass-linewidth radius relationship for the
molecular clouds in M33.  In both plots, the overlaid grey dots
represent the molecular clouds from a Milky Way catalog generated by
merging S87 and \citet{hc01}.  The power-law fits give $\Delta V
\propto r^{0.45 \pm 0.02}$ and $M \propto \Delta V^{4.2 \pm 0.3}$.
The radius measurements for the Milky Way have been recalculated to
match the methods discussed in \S \ref{cloudprops}, though a slight
offset between the data sets persists.}
\end{center}
\end{figure*}

The results of Paper I show that the mass spectrum for GMCs in M33
differs from that found in the Milky Way: $dN/dM \propto M^{-2.6}$ for
M33 vs. $dN/dM \propto M^{-1.5}$ for the Milky Way (S87).  Despite
differences in the mass distribution, clouds in M33 show the same
power law relationships between size, mass and linewidth as are seen
in the Milky Way (Larson's Laws, after \citet{larson}, see also S87).
We plot the size--linewidth and the mass--linewidth relationship for
the clouds in M33 in Figure \ref{larson} along with the data from the
Milky Way \citep{srby87, hc01}.  Where feasible, the cloud properties
of the Milky Way data have been recomputed using the methods in
\S\ref{cloudprops} to ensure a common basis for comparison.  Fitting
power-laws to the two relationships shows $\Delta V \propto r^{0.45
\pm 0.02}$ and $M\propto \Delta V^{4.2\pm 0.3}$.  The fits are
consistent with the relations seen in the Milky Way (S87) in both
normalization and scaling.  However, cloud sizes in M33 are
systematically smaller by $\sim 30$\% in M33 compared to S87, which is
most likely due to differences in the size measurement technique.

\begin{figure}
\begin{center}
\plotone{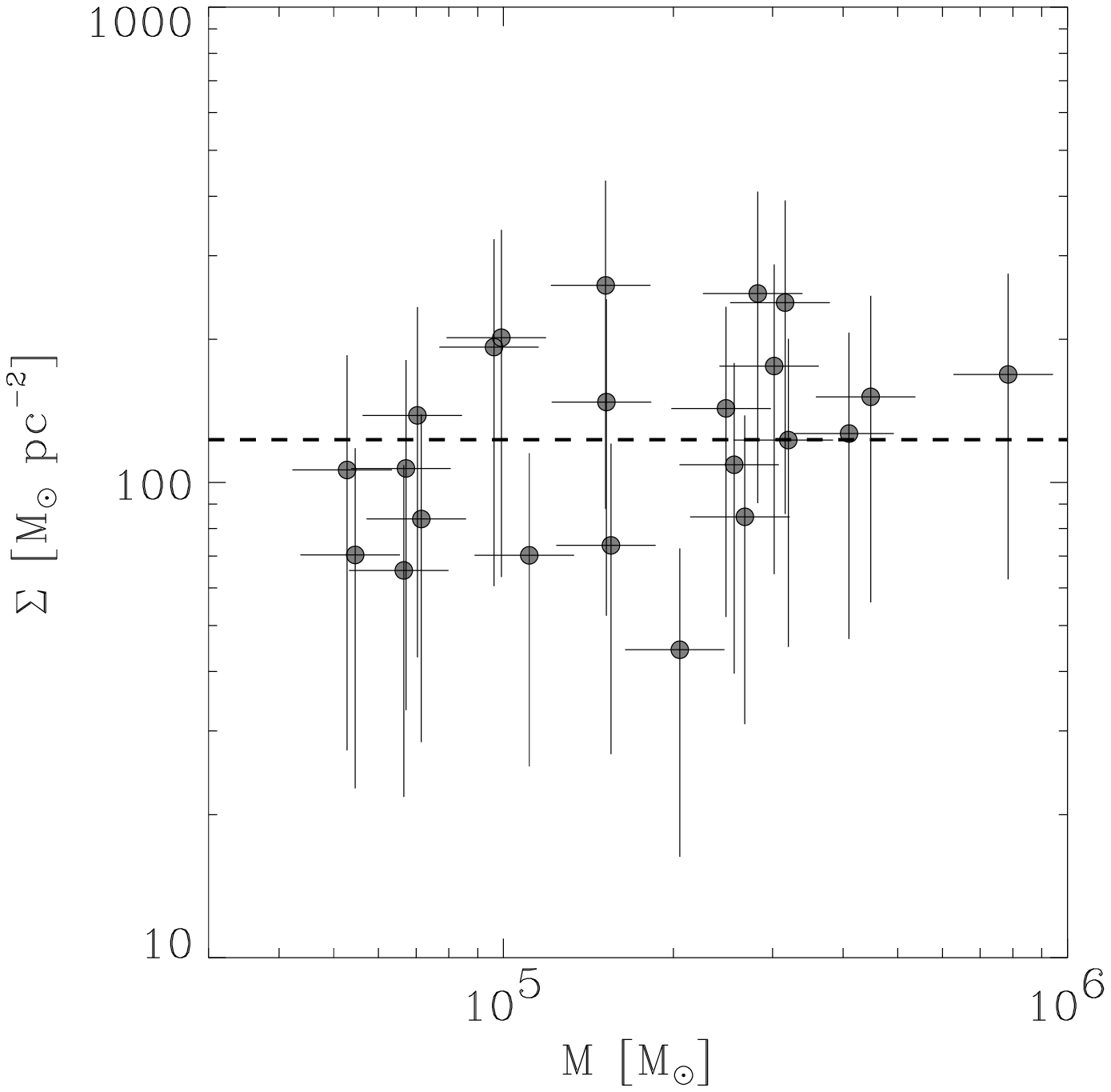} \figcaption{\label{surfdens} Plot of molecular
surface density averaged over GMC as a function of mass for 23
well-resolved, round clouds.  The surface density is calculated using
the deconvolved radius $r$ and the luminous mass $M_{\mathrm{CO}}$.
There is no significant scaling with mass.  Similar plots as a
function of cloud radius and galactic radius also show no scaling.
The mean surface density (dashed line) is $120\ M_{\odot}$ pc$^{-2}$
with a scatter of 60\%.}
\end{center}
\end{figure}

Figure \ref{surfdens} shows the mean surface mass density within
molecular clouds as a function of cloud mass.  This figure shows no
significant scaling of surface density with mass.  Similar plots of
surface density as a function of cloud radius and galactic radius also
show no scaling.  The slopes of linear fits between the mass density
and cloud mass or cloud radius are consistent with zero.  In addition,
two statistical tests of correlation, the Spearman rank order and the
Kendall $\tau$ tests show no significant correlation between mass,
radius and surface density \citep[see][]{numrec}.  We conclude that
GMCs have a surface mass density of $120\ M_\odot\ \mbox{pc}^{-2}$
($2N(\mbox{H}_2)=6\times 10^{21}\mbox{ cm}^{-2}$) with a dispersion of
60\%, which does not vary with cloud mass or radius.  This constant
mass density is also seen for the GMCs in the Milky Way \citep{psp3}.

The consistency of Larson's laws between the Milky Way and M33 shows
that the macroscopic properties of a GMC ($r$, $\Delta V$,
$N(\mbox{H}_2)$) are the same in both galaxies for a given cloud mass.
Since these macroscopic properties follow similar trends, it is
reasonable to expect that the internal properties of GMCs depend only
on the mass of the cloud.  The GMCs throughout the Local Group also
follow Larson's Laws; the LMC \citep{nanten}, Andromeda \citep{sheth},
and dwarf elliptical galaxies \citep{young,young2} all show some
evidence that cloud properties are set by cloud mass.  Some unifying
mechanism must establish the macroscopic properties of the cloud
solely in terms of the cloud mass.  Since turbulence could
provide a relationship between size and linewidth, the balance between
turbulent support and self-gravity would naturally set the macroscopic
properies of a cloud solely in terms of its mass \citep{larson,eg-vt}.

\subsection{The CO-to-H$_2$ Conversion Factor \label{xfactor}}

\begin{figure}
\begin{center}
\plotone{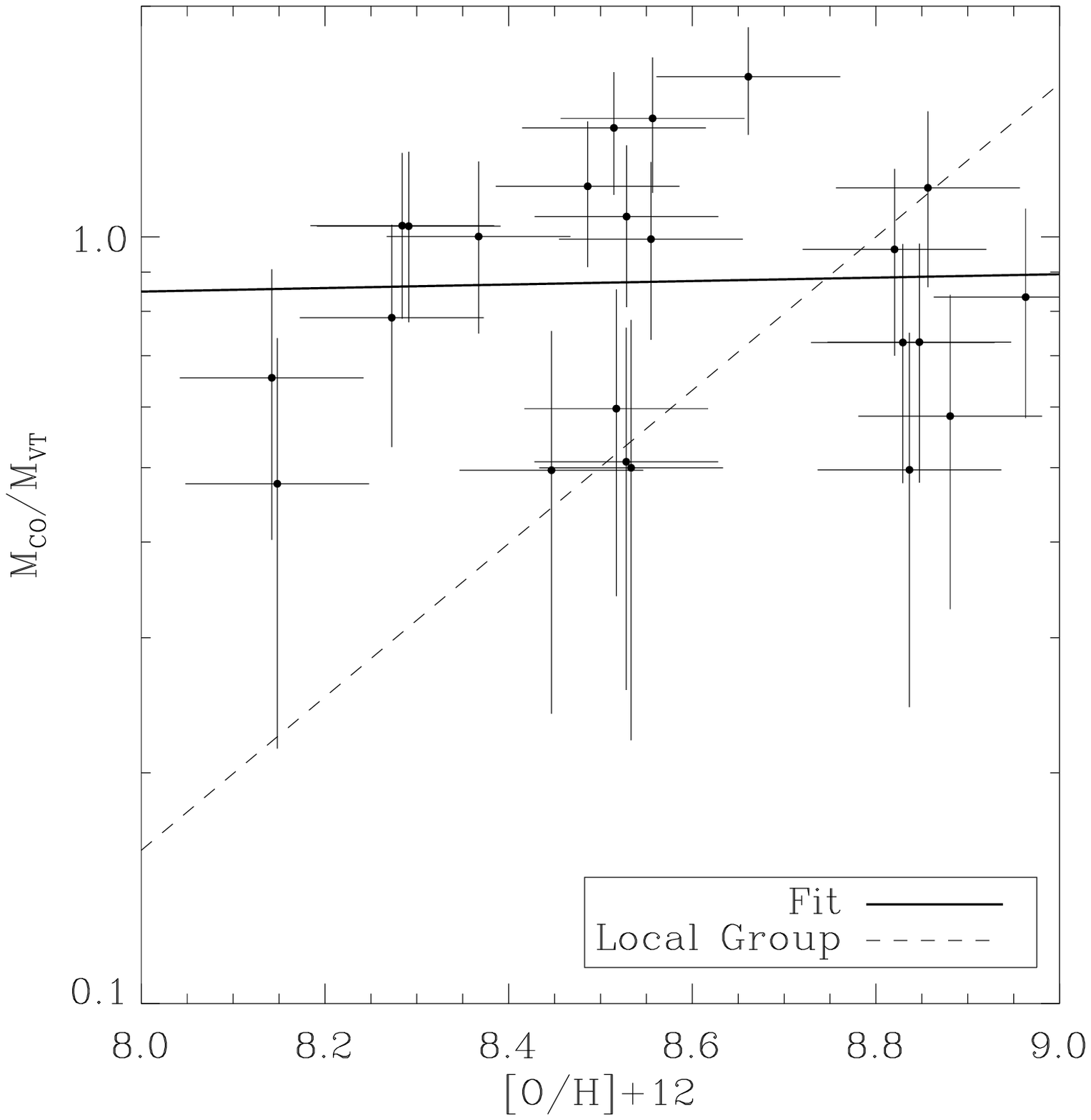} \figcaption{\label{xfac2} Variation of
$M_{\mathrm{CO}}/M_{\mathrm{VT}}$ as a function of metallicity for 23
well-resolved, round clouds in M33.  A linear regression shows no
significant effect from changing metallicities, shown by the solid
line.  The dashed line plots the trend from \citet{xfac96} summarizing
similar measurements throughout the Local Group.}
\end{center}
\end{figure}

In M33, we are presented with a unique opportunity to study the
effects of galactic environment on the CO-to-H$_2$ conversion factor
($X$) since there are no systematic differences due to inhomogeneous
observation techniques or multiple galaxies.  Many studies report a
variation in the $X$ factor as a function of metallicity {\it e.g.}
\citet{xfac96} and references therein.  There is no evidence for such
a trend in M33.  In Figure \ref{xfac2}, we plot the ratio of the
luminous to virial mass $M_{\mathrm{CO}}/M_{\mathrm{VT}}$ as a
function of metallicity for the 23 well-resolved ($r > 10$ pc), round
($f > 0.5$) GMCs in our study.  The metallicity at each point is
obtained using the position of the cloud and the metallicity gradient
($d\mathrm{[O/H]}/dR$) of \citet{hh95}.  Fitting a line to the data
gives:
\[\frac{M_{\mathrm{CO}}}{M_{\mathrm{VT}}} = (0.03 \pm 0.23)
\left([\mathrm{O/H}]-[\mathrm{O/H}]_{\odot}\right)+(0.88 \pm 0.09)\]
accounting for errors in both the metallicity (0.1 dex) and the mass
ratio \citep{numrec}.  The value of $\widetilde{\chi^2}$ for the fit
is 0.9.  The value of $\widetilde{\chi^2}$ using the relationship of
\citet[][dotted line]{xfac96} as a fit gives 3.31, showing poor
agreement ($P=10^{-5}$) with their quoted trend.  The slope of our fit
is consistent with {\it no variation in the conversion factor over a
range of 0.8 dex in metallicity} (a factor of 6).  \citet{bw95} note
that the ratio between virial and CO masses does not change
systematically with radius in the Milky Way.  In \S\ref{mass_sec}, we
argue that interferometer observations overestimate the ratio
$M_{\mathrm{CO}}/M_{\mathrm{VT}}$ by {\it at most} a factor of 1.5.
This cannot account for the discrepancy with the \citet{xfac96} data,
which we attribute to the difficulties in synthesizing a homogeneous
data set from multiple observational studies using a variety of
analysis techniques.

The studied GMCs range in galactocentric radius from 170 pc to 4000
pc, and there are significant changes in galactic properties over this
range.  The interstellar radiation field changes value by nearly an
order of magnitude over this range in galactic radius \citep{isrf}.
Other variations include changes in the cosmic ray flux and the
midplane pressure of the gas.  Such robustness of the $X$ factor as a
GMC mass measure is observed in the Milky Way (S87), but variations
among galaxies may still produce different values in the conversion
factor.  For example, the $X$ factor may change for GMCs in even lower
metallicity systems such as the SMC that have an average metallicity
below the minimum 8.1 observed in M33.

\subsection{Implications for Cloud Formation}
\label{finis}

This study of M33 suggests several attributes of the molecular cloud
formation process.  (1) Molecular clouds form directly from atomic
hydrogen rather than the agglomeration of small molecular clouds.  (2)
The formation process makes both low mass and high mass molecular
clouds in regions of cloud formation. (3) The formation events are
local, {\it i.e.} length scales less than 500 pc.  (4) The progenitor
gas has its initial angular momentum dissipated by MHD effects. (5)
The similarity of molecular clouds across many environments implies
that the molecular mass distribution sets the properties of molecular
clouds and star formation.

Our results complement those of Paper I where it is argued that
molecular clouds form directly from atomic hydrogen.  In that paper,
the striking correspondence between molecular clouds and dense atomic
filaments strongly suggests cloud formation from atomic gas.  This
work strengthens that suggestion since we find no evidence for
diffuse molecular gas that could serve as a precursor to GMCs
(\S\ref{fluxrec}).  Additionally, the atomic gas is significantly more
massive than the molecular component on a global as well as a local
($\sim 100$ pc) scale. If diffuse {\it molecular} gas were the
precursor material, the angular momentum problem would be greatly
exacerbated, scaling predicted angular momenta up by a factor of
$\Sigma_{\mathrm{HI}}/\Sigma_{\mathrm{H}_2,diffuse} > 20$.  We also
find that the flux not recovered in Paper I survey can be accounted
for by halos of molecular gas around the detected GMCs, comprised in
part by low mass molecular clouds (\S\ref{fluxrec}).  These low mass
clouds may form in the same event as the high mass cloud at the
center of the interferometer field.  Such low mass molecular clouds
would have easily been detectable in the off fields of the 12-m
observation and may be restricted to regions where high mass
molecular clouds also form.

The position angle correlations in \S\ref{pacorrs} suggest that the
formation of molecular clouds occurs on length scales less than 500
pc.  A common formation event could produce aligned molecular clouds,
analogous to the formation of the solar system producing aligned
rotation in the planets.  Correlations in the orientation of angular
momentum can be predicted in simulations of turbulent cloud formation
that include realistic models of galactic shear.  Such predictions
should show the observed degree of alignment with the galaxy and
correlations among the GMCs.

The angular momentum problem can be solved by appealing to MHD
braking.  Forming the molecular clouds from a narrow range of radii
requires a long formation time.  Formation mechanisms that accumulate
gas supersonically cannot have any angular momentum dissipated by
Alfv\'en waves since the accumulation time is smaller than crossing
times for magnetosonic waves.  In contrast, the magnetic field is
sufficient to brake forming molecular clouds and braking occurs over
timescales comparable to those of instabilities that could potentially
form molecular clouds (\S\ref{mhdsec}).

Molecular clouds appear to have their macroscopic properties ($R$,
$\Delta V$) set by their masses (\S\ref{comparison}).  In addition,
there appears to be a constant star formation rate per unit molecular
mass within galaxies \citep{wb02}.  If this is true, then the
properties of star formation on a galactic scale depend on the amount
of atomic gas converted into molecular clouds and little else.
Although there are similarities in the properties of molecular clouds
for a fixed mass, there is good evidence that the mass distribution of
molecular gas into individual clouds varies among galaxies.  Different
mass spectra may be the hallmark of multiple formation mechanisms
converting atomic into molecular gas.  The steep mass spectrum for the
molecular clouds in M33 (Paper I) represents the {\it only} difference
in the molecular cloud population between M33 and the Milky Way.  The
mass distribution of molecular gas is thus the dominant factor
controlling star formation on a local and galactic scale.

\section{Conclusions}

We have presented high resolution follow-up observations to a survey
of GMCs in M33 made by \citet{epb02}, using the BIMA array and the UASO
12-m millimeter wave telescope.  A total of 45 individual GMCs were
detected.  The UASO 12-m observations measured the total flux in 18
fields, providing information on the flux lost by the interferometer
and the presence of a diffuse component of molecular gas.  From our
observations, we made the following conclusions.

1. There is no evidence for a diffuse molecular component traced by CO
spanning the disk of M33.  We place a $3\sigma$ surface mass density
limit of $\Sigma_{\mathrm{H}_2} < 0.3\ M_{\odot}$ pc$^{-2}$ on the
presence of such a component.  This is significantly less than the
typical surface density of atomic hydrogen, 7 $M_{\odot}$ pc$^{-2}$.

2. Most of the CO flux in the galaxy is associated with GMCs with
smaller clouds clustered around the large cloud.  Low mass clouds may
appear in other parts of the galaxy but are not ubiquitous.

3. The velocity gradients of the molecular clouds are approximately
linear. The magnitudes of the gradients are comparable to those found
in the Milky Way.  If the gradients are due to rotation, the rotation
period is significantly longer than a cloud lifetime (for
$\tau_{cloud} \leq 30$ Myr).

4. GMCs show significantly smaller angular momenta than are predicted
by simple formation theories.  The discrepancy is, on average, a
factor of 5 and ranges up to two orders of magnitude.  If the velocity
gradients are due to turbulent motions, the discrepancy widens by at
least a factor of 2.  Both the Toomre and Parker instabilities predict
angular momentum values that are discrepant from observations.

5. The specific angular momentum is related to mass as $j(M) \propto
M^{0.6 \pm 0.1}$.  This value is consistent with the derived value for
the Galaxy of $j \propto M^{0.7}$ \citep{p99}.  There is no
significant variation of the specific angular momentum with
galactocentric radius, though such variations are predicted by all the
large scale accumulation theories.

6. The clouds appear somewhat aligned with the rotation of the galaxy,
though a random distribution of position angles is not strongly
excluded.  If the velocity gradients are from rotation, only $\sim
60\%$ of the clouds are prograde rotators.

7. The projected velocity gradients of neighboring, high-mass clouds
are preferentially aligned.  This correlation vanishes for separations
larger than 500 pc.  This alignment may be a signature of the large
scale mechanisms that dictate the formation of molecular clouds and
should be seen in turbulent cloud formation models.

8. The observed molecular clouds are similar to those found in the
Milky Way.  We find a size--linewidth relationship of $\Delta V
\propto r^{0.45 \pm 0.02}$ and a mass--line width relationship of $M
\propto \Delta V^{4.2 \pm 0.3}$.  Both of these relationships are
indistinguishable from those found in the Milky Way.  There is also no
detectable variation in the column density of the molecular clouds
with the mass or radius of the clouds.

9. Equivalent virial masses and luminous masses imply that the $X$
factor for M33 is equal to $2 \times 10^{20} \mbox{ H}_{2} \mbox{
cm}^{-2}/(\mbox{K km s}^{-1})$.  There is no significant variation of
the $X$ factor with metallicity over a range of 0.8 dex.

10. These observations support a model for cloud formation in M33
using atomic gas as the progenitor material for molecular clouds.  The
formation process accumulates the atomic gas over a small distance ($<
500$ pc) with significant braking by magnetic fields.

\acknowledgements We thank an anonymous referee whose comments
improved the paper.  We gratefully acknowledge the help of Edvige
Corbelli for the \ion{H}{1} surface density and rotation curve data
used in her paper.  We thank Adam Leroy for extremely useful
discussions of data reduction and analysis. Our discussions with Ellen
Zweibel regarding MHD braking during cloud formation have proven
especially fruitful.  Observations at the UASO 12-m were aided by
great help from the operator team consisting of Sean Keel, Jon
Carlsen, John Downey and headed by Paul Hart.  We are also grateful to
Tom Dame who provided us with swift checks on our data using the CfA
1.2 m telescope.  ER's work is supported in part by a NSF Graduate
Fellowship and was completed with great encouragement from Elizabeth
Tan.  This research has been made possible by extensive use of NASA's
Astrophysics Data System (ADS) and the NASA/IPAC Extragalactic
Database (NED).

\appendix
\section{Imparted Angular Momentum}
\label{angder}
Consider the progenitor mass for a molecular cloud distributed in
galactocentric radius with a linear mass density $\lambda(R)$.  The
specific angular momentum of this material is 
\[ j =\frac{1}{M}\int \lambda(R) V(R) R\ dR .\]
If the cloud's final distance from the center of the galaxy is $R_c$,
then we define $r\equiv R-R_c$ and perform a Taylor expansion of the
velocity curve around $R_c$ to get
\[ j = \frac{1}{M} \int \lambda(r) \left[V_0 + 
r  D(R_c)\right] \left(r+R_c\right) dr\]
where
\[ D(R_c) \equiv\left.\frac{1}{R}\frac{d}{dR}(VR)\right|_{R=R_c}.\]
Multiplying out the integral gives,
\[ j= \frac{1}{M} \left[M V_0 R_c + \left[R_c D(R_c)+ V_0 \right]
\int \lambda(r) r\ dr + D(R_c) \int \lambda(r) r^2\ dr\right]\] where
the first term represents the angular momentum of a particle
orbiting the center of the galaxy and the other two terms are the spin
angular momentum of the cloud, the measured quantity in these
observations.  In the absence of external forces, the cloud will
collapse to the center of mass, which implies $\int \lambda(r) r = 0$.
In this case, the imparted angular momentum is
\[ j_{gal} = V_0 R_c + \eta  D(R_c) \Delta R^2 \]
where the matter is accumulated from a region of width $2\Delta R$.
We define a constant $\eta$ set by the moment of the mass
distribution in terms of the dimensionless parameter $u=r/\Delta R$:
\begin{equation}
	\eta=\int_{-1}^{1} \lambda\left(u\right)
	u^2 d u.
\end{equation}
For collapse in a circular region of radius $\Delta R$ around $R_c$
with uniform surface mass density, $\eta = 1/4$.

\end{document}